\documentclass[preprint,12pt]{elsarticle}
\usepackage{amsfonts}
\usepackage{graphics}
\usepackage[%nohead,%nofoot
letterpaper, top=1in, bottom=1in, left=1in, right=1in]{geometry}
\usepackage{xcolor}
\usepackage{amssymb}
\usepackage{psfrag}
\usepackage{graphicx}
\usepackage[utf8]{inputenc}
\usepackage{bm}
\usepackage{mathtools}
\usepackage{epstopdf}
\usepackage[colorlinks,linkcolor=black,anchorcolor=black,citecolor=black]{hyperref}
\usepackage{cleveref}
\usepackage{setspace}
\usepackage{amsmath,amssymb,bbm}
\usepackage[utf8]{inputenc}
\usepackage{url}
\usepackage{color}

\newcommand{\be}{\begin{equation}}
\newcommand{\ee}{\end{equation}}

\newtheorem{assum}{Assumption}

\newtheorem{rem}{Remark}

\newtheorem{scheme}{Scheme}

\newproof{proof}{Proof}

\newcommand\undermat[2]{%
  \makebox[0pt][l]{$\smash{\underbrace{\phantom{%
    \begin{matrix}#2\end{matrix}}}_{\text{$#1$}}}$}#2}

\begin{document}
\begin{frontmatter}

\title{
A Feynman-Kac based numerical method for the exit time probability of a class of transport problems
}
%
%
%\thanks{This material is based upon work supported in part by the U.S.~Air Force of Scientific Research under grant numbers FA9550-11-1-0149 and 1854-V521-12; by the U.S.~Department of Energy, Office of Science, Office of Advanced Scientific Computing Research, Applied Mathematics program under contract numbers ERKJ259, and ERKJE45; by the National Natural Science Foundations of China under grant numbers 91130003 and 11171189; by Natural Science Foundation of Shandong Province under grant number ZR2011AZ002;
%and by the Laboratory Directed Research and Development program at the Oak Ridge National Laboratory, which is operated by UT-Battelle, LLC, for the U.S.~Department of Energy under Contract DE-AC05-00OR22725.}

\author[ORNL2]{Minglei Yang}
\author[ORNL1]{Guannan Zhang}
\author[ORNL2]{Diego del-Castillo-Negrete}
\author[ORNL1]{Miroslav Stoyanov}
\address[ORNL2]{Fusion Energy Division, Oak Ridge National Laboratory, Oak Ridge, TN.}
\address[ORNL1]{Computer Science and Mathematics Division, Oak Ridge National Laboratory, Oak Ridge, TN.}

%\author{G.~Zhang\thanks{Department of Computational and Applied Mathematics, Oak Ridge National Laboratory, 
%Oak Ridge, TN 37831({\tt zhangg@ornl.gov}).}
%        \and W.~Zhao\thanks{School of Mathematics, Shandong University, Jinan 250100, China ({\tt wdzhao@sdu.edu.cn}).}
%         \and C.~G.~Webster\thanks{Department of Computational and Applied Mathematics, Oak Ridge National Laboratory, 
%Oak Ridge, TN 37831({\tt webstercg@ornl.gov}).}
%        \and M.~Gunzburger\thanks{Department of Scientific Computing, Florida State University, Tallahassee, FL 32306({\tt gunzburg@fsu.edu}).}
%}
%\maketitle
%
%\pagestyle{myheadings}
%\markboth{G.~Zhang, W.~Zhao, C.~G.~Webster and M.~Gunzburger}
%         {Numerical Solution of BSDEs for Nonlocal Diffusion Equations}

\begin{abstract}
The exit time probability, which gives the likelihood that an initial condition leaves a prescribed region of the phase space of a dynamical system
at, or before, a given time, is arguably one of the most natural and important transport problems. Here we present an accurate and efficient numerical method for computing this probability for systems described by non-autonomous (time-dependent) stochastic differential equations (SDEs) or their equivalent Fokker-Planck partial differential equations. The method is based on the direct approximation of the Feynman-Kac formula that establishes a link between the adjoint Fokker-Planck equation and the forward SDE. The Feynman-Kac formula is approximated using the Gauss-Hermite quadrature rules and piecewise cubic Hermite interpolating polynomials, and a GPU accelerated matrix representation is used to compute the entire time evolution of the exit time probability using a single pass of the algorithm. The method is unconditionally stable, exhibits second order convergence in space, first order convergence in time, and it is straightforward to parallelize. Applications are presented to the advection diffusion of a passive tracer in a fluid flow exhibiting chaotic advection, and to the runaway acceleration of electrons in a plasma in the presence of an electric field, collisions, and radiation damping. Benchmarks against analytical solutions as well as comparisons with explicit and implicit finite difference standard methods for the adjoint Fokker-Planck equation are presented. 
\end{abstract}


%\begin{AMS}
%34A08, 45A05, 45K05, 60G22, 60G52, 60G57, 65C30, 65M15, 65H35, 60H10
%\end{AMS}

%\cortext[cor]{Corresponding author}
\tnotetext[fn1]{{\bf Notice}:  This manuscript has been authored by UT-Battelle, LLC, under contract DE-AC05-00OR22725 with the US Department of Energy (DOE). The US government retains and the publisher, by accepting the article for publication, acknowledges that the US government retains a nonexclusive, paid-up, irrevocable, worldwide license to publish or reproduce the published form of this manuscript, or allow others to do so, for US government purposes. DOE will provide public access to these results of federally sponsored research in accordance with the DOE Public Access Plan.}

%\tnotetext[fn1]{This material is based upon work supported in part by the U.S.~Department of Energy, Office of Science, Offices of Advanced Scientific Computing Research and Fusion Energy Science, and by the Laboratory Directed Research and Development program at the Oak Ridge National Laboratory, which is operated by UT-Battelle, LLC, for the U.S.~Department of Energy under Contract DE-AC05-00OR22725.}
\end{frontmatter}

%%%%%%%%%%%%%%%%%%%%%%%%%%%%%%%%%%%%%%%%%%%%%%
%
%                    Section 2 nonlocal diffusion models and backward SDEs
% we already proposed a probabilistic numerical schemes for approximation of the runaway probability of electrons of time-independent problem \cite{}.
%%%%%%%%%%%%%%%%%%%%%%%%%%%%%%%%%%%%%%%%%%%%%%
\section{Introduction}

The study of transport is a top priority in science and engineering, as well as a 
source of applied mathematics and computational challenges. Some examples, among many others,  include the quantification of the dispersal of pollutants in the atmosphere and the oceans, the design  of efficient mixing protocols in chemical and mechanical engineering, and the study of heat and particle transport in magnetically confined plasmas in controlled nuclear fusion \cite{bagnold1966approach,bird2002transport,rubin1983transport}.
The computation of the probability distribution of the first exit time, denoted by $P(t,{x})$, is arguably one of the most natural and important transport problems \cite{gobet2000weak,baldi1995exact,gobet2004exact}. That is the probability of a particle, with an initial condition ${x} \in \mathbb{R}^d$, to meet, for the first time, a specific target, e.g., reaching, or leaving, a region of the space before a predefined time $t$. The existence of probability distribution of the first exit time has been investigated in Refs.~\cite{strassen1967almost, ferebee1982tangent}. One can find integral equations that can be used to compute this disribution numerically in Refs.~\cite{durbin1985first,park1976evaluations,peskir2002integral,salminen1988first}.
Asymptotic properties of the density are studied in Refs.~\cite{baldi1995exact,ferebee1983asymptotic, uchiyama1980brownian}.
Throughout this paper we restrict attention to Brownian motions, that is stochastic dynamical systems driven by a Wiener process  which underlines the modeling of  local transport in fluids and plasmas.
% The computation of the exit time,  is arguably one the most natural and important transport problems \cite{der_exit:martin,hand_sto,trans:1,case1960elementary}.
% The mean exit time was studied in Refs.~\cite{tuckwell1984first,higham2013mean,buchmann2003computing,bernal2015comparison} that gave a probabilistic representation of the solution of Dirichlet problems \cite{buchmann2006exit,buchmann2003solving}.
% In simple terms, the question is how long does it take for a given initial condition to leave, or reach, a prescribed region of space. 

In Ref.~\cite{zhang2017backward}, we presented a preliminary version of the method proposed here in the context of  the ``runaway acceleration'' problem of electrons in magnetically confined fusion plasma for \emph{time-independent} model parameters. In the present paper, we extend the method to time-dependent parameters, and most importantly to the computation of the full time evolution of the exit probability $P(t,{x})$. Specifically, we present a method for the accurate approximation of $P(t,{x})$ for any $t$ in an interval $[0, T_{\max}]$ (rather than a fixed $t$) in dynamic non-autonomous (i.e. with explicit time-dependent parameters) systems. The crux of the proposed new method, and the main difference with Ref.~\cite{zhang2017backward}, is the computation of  $P(t, x)$ in $[0,T_{\max}]$ with a single pass of the Feynman-Kac formula as  described in  Algorithm 1. An advantage of using the Feynman-Kac representation is that the resulting numerical scheme is \emph{explicit} and \emph{stable}. 
Implicit PDE solvers also have good stability properties but require solving the PDE $N$ times in the non-autonomous case. When using explicit PDE solvers, one can combine $N$ PDE solves into one. But, to guarantee stability the time step needs to be significantly reduced to guarantee stability. In summary,the proposed method significantly reduces the computational complexity of the full dynamics of the exit probability for non-autonomous problems, compared to both  Ref.~\cite{zhang2017backward} and existing PDE solvers. 
Additional contributions of this work, beyond Ref.~\cite{zhang2017backward}, include (i) illustrating the applicability of Feynman-Kac based scheme for 3D runaway electron problem in dynamic scenarios and fluid advection-diffusion problems in the presence of chaotic advection; and (ii) developing and demonstrating a GPU accelerated implementation, critical in solving large-scale problems.

These extensions, accompanied by significant mathematical and computational challenges per se, are motivated by critical physics and engineering application needs. For example, the  modeling of  runaway acceleration of electrons in realistic plasma physics systems requires the incorporation of high-dimensional (up to 6-dimensions) transport effect resulting from adding spatial transport and full-orbit dynamics in configuration space  to the  previously studied (see e.g., Ref.~\cite{wang2016multi,del2018numerical,carbajal2017space}) 2-dimensional transport in velocity space.  Applications also underscore the need to extend the models to full time dependence. For example, plasmas and fluids  are typically far-from-equilibrium and realistic modeling in dynamic scenarios requires the incorporation of time dependent parameters in the SDEs describing these systems. The evolution of these parameters can be given  a priori or determined through a self-consistently coupling of the SDEs with another mathematical model, e.g., a  system of partial differential equations (PDEs) describing the dynamics of the parameters. 
The results of this paper offer a first step to meet these challenges, and as illustration we apply the method to a time-dependent runaway electron problem in 3D (i.e., including spatial transport) and to a fluid mixing problem involving chaotic advection on a time dependent cellular incompressible flow.

A standard numerical method for computing the exit probability is Monte Carlo (MC) ( e.g., \cite{buchmann2003computing,gobet2000weak,bernal2015comparison}), which simulates the subsequent positions of a path by time-stepping schemes  and perform the statistics of the exit condition on a large ensemble of $N$-sample paths for each initial condition of interest \cite{buchmann2005simulation,lansky1994first}. 
The main drawback of the MC method is the unfavorable scaling, $\mathcal{O}( 1/\sqrt{N})$, of the statistical sampling error with the number of samples, $N$. 
% Although in recent works, e.g.  \cite{buchmann2003computing,gobet2000weak,bernal2015comparison}, direct integration algorithms for SDEs achieving order 
% $\Delta t$ accuracy have been proposed, the direct method still faces the 
% $\mathcal{O}(1/\sqrt{N})$ low convergence. 
Among the PDE based numerical approaches, the exit probability can be computed by first solving the forward Fokker-Planck equation \cite{risken1996fokker,yang2018numerical,spencer1993numerical,park1996fokker} associated with the SDE, and then convolve the probability density function of SDE solution with the exit condition.  

% first one is to obtain the probability density function of the path by solving the forward Fokker-Planck equation associated with the stochastic differential equations through some deterministic approaches 

% as well as the overhead in the numerical integration for each initial condition, which can be significant for methods more sophisticate than Euler stepping.

Both the MC method and the PDE method for the forward Fokker-Planck equation depends on the initial condition. When studying the dependence of the probability of the exit time on large families of initial conditions, which is a problem of interest in scientific applications, these methods are not efficient since they require running the algorithm repeatedly. As it will be explained below, our method overcomes this limitation because the Feynman-Kac formula provides a type of Green's function representation for the exit time probability, from which the exit probability can be obtained by convolution with the initial condition. 

An alternative approach is to solve the backward (adjoint) Fokker-Planck equation associated with the SDE \cite{mark1985functional,brannan2001escape,yoshioka2012partial,patie2008first}. This method reduces the computation of the probability of exit time to the solution of a terminal-value problem of a parabolic PDE. Although superior to MC sampling, the adjoint methods share the limitations of PDE solvers including numerical instability, poor scalability with dimensions, and algorithmic challenges in parallelization. Also, since the adjoint methods are based on the solution of a terminal-value problem, in principle they only provide $P(t,{x})$ for a fixed $t$. That is, computing $P(t,{x})$ for $t \in \{t_1, \, t_2, \, \ldots t_M \}$ requires to solve the adjoint problem $M$-times. 

We propose a different method based on the direct approximation of the Feynman-Kac formula that establishes the connection between the solution of the (backward) adjoint Fokker-Planck equation and the associated forward SDEs.
%Here we generalized and optimized the numerical scheme proposed in Ref.~\cite{Spare:RE} to a matrix structure algorithm that bypasses some of these limitations. 
We express the stochastic representation of the solution of the adjoint equation as mathematical expectations and compute these expectations using Gauss-Hermite quadrature rules and piecewise cubic Hermite interpolating polynomials (PCHIP). Three quadrature points in each dimension are sufficient to guarantee first order, $~\Delta t$, global convergence in time. The shape-preserving PCHIP interpolation guarantees that all the interpolated values are between 0 and 1.  
% To eliminate the error caused by directly approximating the random exit time, 
%  To eliminate the destructive effects of the probability of the random exit time,
To avoid direct approximation of the random exit time,
we exploit the regularity of the exit probability \cite{Yang:2018fd}. Specifically, if the starting point of the state process is far from the boundary, the exit probability decays sub-exponentially, and the high-order PCHIP interpolation compensates the accuracy loss near the boundary. 
Moreover, the proposed algorithm only needs one temporal loop to recover the full dynamics of exit probability, i.e., $P(t,{x})$ for $t \in \{0=t_1 < t_2< \cdots < t_M = T_{\max}\}$, and all propagation information only assemble once at each time step. The main features of the proposed method are summarized as follows:
\vspace{-0.1cm}
\begin{itemize}\itemsep-0.05cm
    \item[(a)] The unconditional stability for any $\Delta t$, first-order convergence with respect to $\Delta t$, and second-order convergence with respect to $\Delta x$;
    \item[(b)] The output of our method can be used to obtain the exit probability of the SDE with different initial conditions by doing simple convolution;
    \item[(c)] Recovering the entire dynamics of the exit probability $P(t,x)$ by only one temporal iteration, and easy implementation on GPUs. 
\end{itemize}
\vspace{-0.1cm}
To our knowledge, our method is the first one that possesses all the three features among existing methods for computing $P(t,x)$. In comparison, the MC methods have feature (c); the PDE solvers for the forward Fokker-Planck equations have feature (a), (c); the implicit PDE solvers for the backward (adjoint) Fokker-Planck equations have feature (a), (b); the explicit solvers for the backward (adjoint) Fokker-Planck equations have feature (b), (c).

The outline of the paper is as follows: In section \ref{prob:set}, we introduce the formal definitions of exit time and exit probability. Moreover, we give mathematical descriptions of the stochastic dynamical system and the adjoint equation. In section \ref{sec:PT}, the fully-discrete probabilistic scheme and the determination of spatial discretization are given. In section \ref{sec:PT2}, we provide an efficient implementation for computing the full dynamics of exit probability. Practical numerical examples are given in section \ref{sec:ex}, which are consistent with theoretical results. Some conclusions are shown in section \ref{sec:clu}.

% \newpage
\section{Problem setting}\label{prob:set}
%Let $(\Omega, \mathcal{F}, \{\mathcal{F}_t\}_{0 \le t \le T_{\max}}, \mathbb{P})$ for $T_{\max}>0$ be a complete probability space 
%with the filtration $\{\mathcal{F}_t\}_{0 \le t \le T_{\max}}$, generated by the $m$-dimensional standard Brownian motion 
% $W_t := (W_t^1, \ldots, W_t^m)^{\top}$. 
Our starting point is the  $d$-dimensional stochastic differential equation
% defined in 
% $(\Omega, \mathcal{F}, \{\mathcal{F}_t\}_{0 \le t \le T_{\max}}, \mathbb{P})$, i.e.,
\begin{equation}\label{e1}
X_t  =   X_0 + \int_0^t {  b}(s, X_s) ds
+  \int_0^t{ \sigma}(s,  X_s) d  W_s \;\; \text{ with } X_0 \in \mathcal{D} \subset \mathbb{R}^d,
\end{equation}
where $W_t := (W_t^1, \ldots, W_t^m)^{\top}$ is an $m$-dimensional standard Brownian motion,
$X_t \in \mathbb{R}^d$. $b: [0,T_{\max}] \times  \mathbb{R}^d \rightarrow \mathbb{R}^d$, $\sigma :[0,T_{\max}] \times \ \mathbb{R}^d \rightarrow \mathbb{R}^{d\times m}$ are globally Lipschitz in $x$ uniformly with respect with respect to $t$, and locally bounded. And  $X_0$ is an 
initial condition in an open bounded domain $\mathcal{D} = (\alpha_1, \beta_1) \times \cdots \times (\alpha_d, \beta_d) \subset \mathbb{R}^d$. The first exit time, $\tau$, is defined as
\begin{equation}\label{e101}
\tau := \inf \big\{ t > 0 \,|\, X_0 \in \mathcal{D}, X_t \not\in \mathcal{D}\big\},
\end{equation}
which is the first time instant when $X_t$ in Eq.~\eqref{e1} exits the domain $\mathcal{D}$.

%\paragraph{The exit probability} 
Our objective is to numerically compute the {\em exit probability}, $P(t,x)$, defined as the probability of the event that the stochastic process in Eq.~\eqref{e1} exits  the domain $\mathcal{D}$ at or before a specific time  $t \in (0, T_{\max})$
\begin{equation}\label{e2}
P(t,x) = \mathbb{P}\left\{ \tau \le t \,|\, X_0 = x \in \mathcal{D} \right\}\; \text{ for any }\; (t,x) \in [0,T_{\max}] \times \overline{\mathcal{D}}.
\end{equation}
Note that $P(t,x)$ in Eq.~\eqref{e2} is independent of the distribution, $P_{\rm init}(X_0)$, of the initial condition, $X_0$.  Once $P(t,x)$ is computed, the exit probability conditional on an initial distribution $P_{\rm init}(X_0)$ is given by the convolution
\[
\mathbb{P}\left\{ \tau \le t \,|\, X_0 \sim P_{\rm init}(X_0) \right\} = \int P(t,x) P_{\rm init}(x) dx.
\]
In this sense, $P(t,x)$ in Eq.~\eqref{e2} can be viewed as a Green's function for the exit probability problem of interest. This feature can save significant computing effort when studying the dynamics of $P(t,x)$ 
as function of different distributions of initial conditions as typically done in applied problems  (e.g., the runaway electron acceleration  problem in Section \ref{e200}).

For a fixed $T \in (0,T_{\max})$, the exit probability $P(T,x)$ in Eq.~\eqref{e2} can be obtained 
from 
\begin{equation}\label{e5}
P(T,x) = u(0, x), 
\end{equation}
where $u(t,x)$ is the solution of the 
parabolic terminal boundary value problem of the 
{\em backward (adjoint) Fokker-Planck equation}
\cite{Schuss:2013th},
\begin{equation}\label{e3}
\begin{aligned}
\frac{\partial u(t,x)}{\partial t}  + \mathcal{L}^{*}(t,x) [u(t,x)] & = 0 \quad \text{ for }\;\; x \in \mathcal{D}, t < T,\\
u(t,x) & = 1  \quad \text{ for }\;\; x \in \mathcal{\partial D}, t<T,\\[4pt]
 u(T,x) & = 0  \quad \text{ for }\;\; x \in \mathcal{D},\\
\end{aligned}
\end{equation}
where the operator $\mathcal{L}^{*}(t,x)$, which is
the adjoint of the forward Fokker-Planck operator associated with $X_t$, is given by
\[
\mathcal{L}^{*}(t,x)[u] := \sum_{i=1}^{d}b_{i}(t,x)\frac{\partial u}{\partial x^{i}}(t,x) + \frac{1}{2}\sum_{i,j=1}^{d}({\sigma}{\sigma}^{\top})_{i,j} \frac{\partial^2 u}{\partial x^{i}x^{j}}(t,x),
\]
with $b_i$ the $i$-th component of the drift $b(t,x)$, $({\sigma}{\sigma}^{\top})_{i,j}$ the $(i,j)$-th entry of the diffusivity tensor ${\sigma}{\sigma}^{\top}$, and $x^i$ the $i$-th component of $x$ in Eq.~\eqref{e1}. 
%The operator $\mathcal{L}^{*}(t,x)$ is the adjoint of the forward Fokker-Planck operator associated with $X_t$.
%, and the PDE in Eq.~\eqref{e3} is referred to as the backward adjoint equation. 
%The exit probability $P(T,x)$ can be represented by
%
%\begin{equation}\label{e5}
%P(T,x) = u(0, x).
%\end{equation}

The full dynamics of $P(t,x)$ for any $(t,x) \in [0,T_{\max}] \times \mathcal{D}$ can then be approximated by solving Eq.~\eqref{e3} $M$-times for the terminal times
\begin{equation}\label{time1}
\mathcal{T} := \{0=T_0<T_1<\cdots<T_{M}=T_{\max}\},
\end{equation}
and then interpolating the solution outside the grid points. 
Although classical PDE solvers based on finite difference or finite element methods can be used to solve Eq.~\eqref{e3}, 
these methods might be inefficient in approximating the entire dynamics of $P(t,x)$ due to time stepping numerical stability issue. 
Using explicit time stepping methods allows to stack the $M$ terminal value problems together in a matrix and push through explicit propagators. However, explicit schemes usually require small time step size to guarantee numerical stability. Implicit time stepping can guarantee  stability. However, implicit methods require solving a linear system at each time step, such that multiple linear systems (e.g., $M$ linear systems at $T_0$) need to be solved independently at each time step. 
As it will be discussed below, the proposed numerical method bypasses these difficulties by using an explicit yet {\em unconditionally stable} time stepping algorithm. 

In applications of interest to plasma physics, as well as in other areas, the parameters in the SDEs, e.g., $\sigma$ and $b$ in Eq~(\ref{e1}), are determined from a self-consistent coupling to another mathematical model governing their evolution.  Although incorporating this type of  couplings using PDE methods is in principle possible, in practice it can be quite difficult and the proposed method can be used to overcome these challenges.

% Hence, we will introduce our numerical scheme for solving $P(T, x)$ at a fixed $T$ in Section \ref{sec:PT}, and extend it to the entire domain $[0, T_{\max}]$ in Section \ref{sec:PT2}.

\section{The proposed Feynman-Kac based numerical method
%Our method for approximating the exit probability
}\label{sec:PT}
%This section provides the details of our method for computing $P(t,x)$ in Eq.~\eqref{e2}. 
The theoretical foundation of our probabilistic method is the Feynman-Kac theory that links the SDE in Eq.~\eqref{e1} to the adjoint equation in Eq.~\eqref{e3}, see for example Refs.~\cite{Pardoux:1992jo,Zhang:2016fb}. 
In Sec.~\ref{sec:time} we focus on the computation of $P(T,x)$ at a fixed time. In Sec.~\ref{sec:PT2} we present an efficient algorithm to compute $P(t,x)$ for $t\in [0, T_{\max}]$ based on the simultaneous solution of $M$ adjoint equations with $M$ different terminal time instants.

% This section focuses on solving the adjoint equation in Eq.~\eqref{e3} for a fixed $T \in [0,T_{\max}]$. The probabilistic representation of $v(t,x)$ and the temporal discretization is given in Section \ref{sec:time}; spatial discretization including a special treatment of the involved random exit time $\tau$ is provided in Section \ref{sec:interp} and \ref{sec:exp}.

\subsection{The numerical scheme for the exit probability at a fixed time}
\label{sec:time}
The first step is to write the probabilistic representation of the solution $u(t,x)$ in Eq.~\eqref{e3}. To this end, we  rewrite the SDE in Eq.~\eqref{e1} in the conditional form
\begin{equation}\label{eq:SDE}
  X_s^{t,  x}  =   x + \int_t^s {  b}(r,   X_r^{t,  x}  ) dr 
+  \int_t^s{  \sigma}(r,  X_r^{t,  x}  ) d  W_r \;\; \text{ for } \; s \ge t,
\end{equation}
where the superscript ${}^{t, x}$ indicates the condition that $X_s^{t,x}$ starts from $(t,x) \in [0,T_N]  \times \mathcal{D}$. Note that $T_N$ is the $N$-th element of the set $\mathcal{T}$ in Eq.~\eqref{time1}.
Accordingly, we can define the \emph{first conditional exit time}
\begin{equation}\label{e22}
\tau_{t, x}  := \inf\{ s > t \,|\, x \in \mathcal{D},  X_s^{t,   x}\not\in \mathcal{D}\}.
\end{equation}
The probabilistic representation of $u(t,x)$ in Eq.~\eqref{e3} is given by the Feynman-Kac formula \cite{pardoux1998backward,freidlin2016functional}
\begin{equation}\label{e23}
u(t,x) = \mathbb{E}\left[u\left(s\wedge \tau_{t,x}, X_{s\wedge \tau_{t,x}}^{t,  x} \right) \right],
\end{equation}
where $s \wedge \tau_{t,x} := \min(s, \tau_{t,x})$ denotes the minimum of $\tau_{t,x}$ and $s$ and $X^{t,x}_{s \wedge \tau_{t,x}}$ is defined based on Eq.~\eqref{eq:SDE}. Instead of solving the PDE in Eq.~\eqref{e3}, our approach is based on the direct computation of $u(t,x)$ by discretizing Eq.~\eqref{e23}.
The temporal and spatial approximations used to accomplish this are introduced in Section \ref{sec:time1} and Section \ref{sec:interp}, respectively. The estimation of the mathematical expectation $\mathbb{E}[\cdot]$ is introduced in Section \ref{sec:exp}.

\subsubsection{Temporal approximation of $u(t,x)$}\label{sec:time1}
We introduce a uniform time partition for $[0,T_N]$:
\begin{equation}
\mathcal{T}_N := \{0=t_0<t_1<\cdots<t_{N}=T\}
\end{equation}
with $\Delta t=t_{n+1}-t_n$ for $n = 0,1, \ldots, N$. We assume $\mathcal{T}_N$ is a subset of $\mathcal{T}$ in Eq.~\eqref{time1}, which means $t_n=T_n$ for $n = 0, \ldots, N$. 
We discretize the SDE in Eq.~\eqref{eq:SDE} in the interval $[t_n, t_{n+1}]$ using a forward Euler scheme \cite{Zhao:2012ku,Platen:2010eo}:
\begin{equation}\label{ref-X}
X_{n+1}^{t_n,x} = x + b(t_n,x)\Delta t+ \sigma(t_n,x)\Delta W, 
\end{equation}
where $\Delta W= W_{t_{n+1}}-W_{t_n}$. Substituting Eq.~\eqref{ref-X} into Eq.~\eqref{e23}, we approximate the solution of $u(t, x)$ at $t=t_n$ using
\begin{equation}\label{e24}
u(t_n,x) \approx u^n(x) = \mathbb{E}\left[ u^{n+1}\left(X_{n+1}^{t_n,x}\right)\mathbf{1}_{\{\tau_{t_n,x} > t_{n+1}\}}\right] +  \mathbb{P}\left(\tau_{t_n,x} \le t_{n+1}\right),
\end{equation}
where  $\tau_{t_n, x}$ is defined in Eq.~\eqref{e22}\footnote{The exit time $\tau_{t_n,x}$ in Eq.~\eqref{e22} should be defined by replacing $X_s^{t,x}$ with the Euler discretization, i.e., $X_{s}^{t_n,x} = x + b(t_n,x)(s-t_n)+ \sigma(t_n,x)(W_s-W_{t_n})$ for $s\ge t_n$ in Eq.~\eqref{e22}. We use the same notation without creating confusion.},
$\mathbf{1}_{\{\tau_{t_n,x} > t_{n+1}\}}$ is the indicator function of the event that $X_{s}^{t_n,x}$ does not exit from $\mathcal{D}$ before $t_{n+1}$, and $\mathbb{P}\left(\tau_{t_n,x} \le t_{n+1}\right)$ is the probability that $X_s^{t_n,x}$ exits from $\mathcal{D}$ within $(t_n, t_{n+1}]$.

\subsubsection{Spatial approximation of $u(t,x)$}\label{sec:interp}
To approximate $u^n(x)$ in $\mathcal{D}$ we use piecewise polynomial interpolation. We first introduce a Cartesian mesh $\mathcal{S} = \mathcal{S}^1 \times \mathcal{S}^2 \times \cdots \times \mathcal{S}^d$ for the domain $\overline{\mathcal{D}}$, where $\mathcal{S}^i$ for $i = 1, \ldots, d$ is a mesh of the interval $[\alpha_i, \beta_i]$. It should be noted that the partition along each direction may be non-uniform.
For notational simplicity, we use $J$ to denote the total number of grid points in $\mathcal{S}$, and use a scalar $j$ to index all the grid points in $\mathcal{S}$, i.e.,
\[
\mathcal{S} := \{ x_j: \; j = 1, \ldots, J\}.
\]
The general formula for piecewise interpolation using nodal values can be written as
\begin{equation}\label{Lag11}
{u}^{n,J}(x) := \sum_{j = 1}^J u^{n}_j\; {\psi}_{j}(x) + \langle \mathbf{d  u}^n_j\,,{\Phi}_{j}(x)\rangle
, \;\; \text{ for } \; x \in \overline{\mathcal{D}},
\end{equation}
where $u_j^n$ for $j=1, \ldots J$ are the approximate nodal values of $u(t_n,x)$, $\mathbf{d  u}^n_j$ for $j=1, \ldots J$ are gradient at nodal ${x_j}$. ${\psi}_{j}(x),{\Phi}_{j}(x)$ 
are generated by assembly all the basis functions associated with the $j$-th nodal value $u_j^n$ and $\mathbf{d  u}^n_j$. Since $u(t,x)$ is a probability taking values within $[0,1]$, we use the Piecewise Cubic Hermite Interpolating Polynomials (PCHIP) \cite{doi:10.1137/0717021} to implement Eq.~\eqref{Lag11}. A brief description of the used interpolation scheme is given below.

\paragraph{Piecewise cubic Hermite interpolating polynomials} 
For simplicity, we briefly recall the one-dimensional PCHIP interpolation. 
Since we are using a Cartesian grid, the one-dimensional scheme can be easily extended to multi-dimensional grid by tensor product \cite{moler2004numerical}. Taking $[x_j, x_{j+1}]$ as an example, the PCHIP approximation of $u^n(x)$ in $[x_j, x_{j+1}]$ can be written as 
\begin{equation}
    \begin{aligned}
    u^{n,J}(x) =& \frac{3hs^2-2s^3}{h^3}u^n_{j+1} + \frac{h^3-3hs^2+2s^3}{h^3}u^n_j\\
    & + \frac{s^2(s-h)}{h^2}\frac{d u^n}{dx}(x_{j+1}) + \frac{s(s-h)^2}{h^2}\frac{d u^n}{dx}(x_{j}),
    \end{aligned}
\end{equation}
where $s = x-x_j$ and $h = x_{j+1}-x_{j}$. As the derivatives $\frac{d u^n}{dx}(x_{j+1})$ and $\frac{d u^n}{dx}(x_{j})$ are unavailable, approximated derivatives are used instead. When $x_j$ is an interior node, $\frac{d u^n}{dx}(x_{j})$ is approximated by
\begin{equation}\label{e300}
  \frac{d u^n}{dx}(x_{j}) =
    \begin{cases}
      0 & \text{if $\delta_j$ and $\delta_{j-1}$ have opposite signs},\\
      \frac{2}{\frac{1}{\delta_{j-1}} + \frac{1}{\delta_{j}}} & \text{if $\delta_j$ and $\delta_{j-1}$ have same sign},
    \end{cases}       
\end{equation}
where $\delta_{j}$ is the first divided difference given by
$
\delta_j = \frac{u_{j+1}^n-u_{j}^n}{x_{j+1}-x_j}.
$
For the boundary node $x_1$, $\frac{d u^n}{dx}(x_1)$ is defined by substituting $\delta_1$ and $\delta_2$ into Eq.~\eqref{e300}; for the boundary node $x_J$, $\frac{d u^n}{dx}(x_J)$ is defined by substituting $\delta_{J-1}$ and $\delta_{J-2}$ into Eq.~\eqref{e300}. 

% {\bf [NOTE: I do not understand (14), should it be the case that $u^{n,J}(x_{j+1})=u^n_{j+1}$? Also, what exactly are the functions $\Phi_j(x)$ and
% $\psi_j(x)$ in (13)?]}\\

% {\bf [(Eq (14) is the pchip approximation formula in one dimensional domain, and $u^{n,J}(x_{j+1})=u^n_{j+1}$. In other words, there are four freedoms, values and derivatives of two endpoints in one interval $[x_j,x_{j+1}]$, to construct the pchip interpolation formula; For the general case, it is hard to exactly express the basis polynomials $\Phi_j(x)$ and $\psi_j(x)$], it depends on dimensionality and method. In my opinion, we can just mention them as basis polynomial here, we already gave a matrix structure of  $\Phi_j(x)$ and $\psi_j(x)$ below Eq(23)}
% Moreover, the first derivative $P'(x_1)$ is given by $\delta_1$, and $\delta_2$; $P'(x_n)$ is given by $\delta_{n-2}$ and $\delta_{n-1}$. The details are in pchiptx.m \cite{moler2004numerical}.

% Then the temporal discretization scheme in Eq.~\eqref{e24} can be spatially discretized by replacing $u^{n+1}$, $u^n$ with $u^{n+1,J}$, $u^{n,J}$, respectively. 

\subsubsection{Treatment of the probability $\mathbb{P}\left(\tau_{t_n,x} \le t_{n+1}\right)$}\label{sec:exit}
$\mathbb{P}\left(\tau_{t_n,x} \le t_{n+1}\right)$ in Eq.~\eqref{e24} describes the probability that $X_s^{t_n,x}$ exits from $\mathcal{D}$ within $(t_n, t_{n+1}]$. It is challenging to approximate the probability with high-order accuracy for any $x$ in $\mathcal{D}$. In our method, we only need to deal with $\mathbb{P}\left(\tau_{t_n,x} \le t_{n+1}\right)$ with $x$ being the interior nodes of the spatial mesh $\mathcal{S}$. Thus, our idea is to impose an additional condition on $\mathcal{S}$ to guarantee that $\mathbb{P}\left(\tau_{t_n,x} \le t_{n+1}\right)$ is sufficiently small, e.g., on the order of the temporal or spatial approximation errors. 

%
% Then the task becomes to estimate the right-hand side of Eq.~\eqref{e24} at all the interior grid points $x_i \in \mathcal{S} \cap \mathcal{D}$. The accuracy of such estimation depends on how to deal with $\mathbb{P}(\tau_{t_n,x} \le t_{n+1})$. 
It is known that $\mathbb{P}(\tau_{t_n,x} \le t_{n+1}) \rightarrow 1$ as $x \rightarrow \partial \mathcal{D}$. In our previous work \cite{Yang:2018fd}, we proved that if $b$ and $\sigma$ are bounded functions, i.e.,
$|b(t,x)| \le \overline{b}$ and $|\sigma(t,x)| \le \overline{\sigma}$ for $(t,x) \in [0,T_{\max}]\times \mathcal{D}$ 
with $0 \le \overline{b}, \overline{\sigma} \le +\infty$, and $x$ in Eq.~\eqref{ref-X} is sufficiently far from the boundary $\partial \mathcal{D}$ satisfying 
$
dist(x, \partial \mathcal{D}) \sim \mathcal{O}((\Delta t)^{1/2-\varepsilon})
$
 for any given constant $\varepsilon >0$, then for sufficiently small $\Delta t$, it holds that
\begin{equation}\label{stoperr}
\mathbb{P}(\tau_{t_n,x} \le t_{n+1}) \le C (\Delta t)^\varepsilon \exp\left(-\frac{1}{(\Delta t)^{2\varepsilon}}\right),
\end{equation}
where the constant $C>0$ only depends on upper constants $\Bar{b}$, $\bar{\sigma}$.
Even though the estimate in Eq.~\eqref{stoperr} was proved in \cite{Yang:2018fd} for the one-dimensional case, it provides a good insight to deal with $\mathbb{P}\left(\tau_{t_n,x} \le t_{n+1}\right)$. Specifically, we impose the following assumption on the spatial mesh $\mathcal{S}$.
\begin{assum}\label{as1}
For any interior node $x_j \in \mathcal{S} \cap \mathcal{D}$, it holds that
\begin{equation}\label{e301}
dist(x_j, \partial \mathcal{D}) \ge C\sqrt{\Delta t},
\end{equation}
where the constant $C>0$ only depends on upper constants $\Bar{b}$, $\bar{\sigma}$, $dist(x_j, \partial \mathcal{D})$ is the smallest Euclidean distance between $x_j$ and $\partial\mathcal{D}$ for $x_j \in \mathcal{S} \cap \mathcal{D}$.
\end{assum}

Note that this assumption is much weaker than a condition on the maximum spatial mesh size, because it only applies to the closest layer of the interior nodes to the boundary. In other words, it provides sufficient flexibility to use non-uniform spatial meshes to handle irregular behaviors (e.g., the sharp transition layer of the RE problem in Section \ref{e200}). How to realize Assumption \ref{as1} in practice will be discussed in Remark \ref{rem1}.

When $\mathcal{S}$ satisfies Assumption \ref{as1}, the probability
$\mathbb{P}(\tau_{t_n,x} \le t_{n+1}) = 1- \mathbb{P}(\tau_{t_n,x} > t_{n+1})$
% $\mathbb{P}(\tau_{t_n,x} \le t_{n+1}) = 1- \mathbb{P}(\tau_{t_n,x} < t_{n+1})$
is very close to zero, which implies that
\[
|\mathbb{E}\left[ u^{n+1}\left(X_{n+1}^{t_n,x}\right)\right]- \mathbb{E}\left[ u^{n+1}\left(X_{n+1}^{t_n,x}\right)\mathbf{1}_{\{\tau_{t_n,x} > t_{n+1}\}}\right]| \sim \mathcal{O}(\mathbb{P}(\tau_{t_n,x} \le t_{n+1})),
\]
because $u^{n+1}$ takes values from $[0,1]$. 
% This means the indicator function $\mathbf{1}_{\{\tau_{t_n,x} > t_{n+1}\}}$ can also be neglected in Eq.~\eqref{e24}.
This means the indicator function $\mathbf{1}_{\{\tau_{t_n,x} > t_{n+1}\}}$ approaches 1 in Eq.~\eqref{e24}.
Hence, by substituting the spatial approximation $u^{n,J}$, $u^{n+1,J}$ into Eq.~\eqref{e24} and imposing Assumption \ref{as1} on $\mathcal{S}$, Eq.~\eqref{e24} reduces to 
\begin{equation}\label{e30}
u^n_j = \mathbb{E}\big[ u^{n+1,J}\big(X_{n+1}^{t_n,x_j}\big)\big] \;\; \text{ for } x_j \in \mathcal{S}\cap\mathcal{D},
\end{equation}
and $u^{n,J}(x)$ can be obtained by interpolating the nodal values $u^n_j$ using Eq.~\eqref{Lag11}.

% The key idea is to eliminate the destructive effect of $\mathbb{P}(\tau_{t_n,x} \le t_{n+1})$ in the construction of the temporal-spatial discretization scheme by exploiting the estimate in Eq.~\eqref{stoperr}. 
% Specifically, we define the spatial mesh size $\Delta x_i$ such that all quadrature points locate in the domain $\overline{\mathcal{D}}$ for each interior grid point $x_i$, the definition of quadrature points and the specific choice of $\Delta x_i$ will be given in Section \ref{sec:exp}. 

% Then the value $v^n(x_i)$ in Eq.~\eqref{e24} can be approximated by

% with the error on the order of $\mathcal{O}((\Delta t)^\varepsilon \exp(-{1}/{(\Delta t)^{2\varepsilon}}))$. Such strategy can avoid the approximation of the exit probability $\mathbb{P}(\tau_{t_n,x} \le t_{n+1})$, but the trade-off is that we need to use higher order interpolation to balance the total error. In this work, since the Euler scheme is used for temporal discretization, the global temporal truncation error will be first order $\mathcal{O}(\Delta t)$. To match that, we need to use cubic interpolation that gives $\mathcal{O}((\Delta x)^4) \approx \mathcal{O}((\Delta t)^{2-4\varepsilon})$ local truncation error and thus $\mathcal{O}((\Delta t)^{1-2\varepsilon})$ global truncation error. Furthermore, we apply pchip interpolation to keep the shape of interpolants since the exit probability's range is $[0,1]$.

\subsubsection{Quadrature for the conditional expectation}\label{sec:exp}
The last piece of the algorithm is a quadrature rule for estimating the conditional expectations for the interior nodes $x_j \in \mathcal{S} \cap \mathcal{D}$ given by the $d$-dimensional integrals
%$\mathbb{E}[ u^{n+1,J}\big(X_{n+1}^{t_n,x_j}\big)]$ 
%The expectation is written as
%
\begin{equation}
\label{expec}
\mathbb{E}\big[u^{n+1,J}(X_{n+1}^{t_n,x_j})\big] = 
\int_{\mathbb{R}^d} u^{n+1,J}\left(x_j+ b(t_n, x_j) \Delta t + \sigma(t_n, x_j)\sqrt{2\Delta t}\, \xi \right) \rho(\xi) d\xi,
\end{equation}
where $\xi := (\xi_1, \ldots, \xi_d)$ follows the  normal distribution with density 
$
\rho(\xi) := \pi^{-d/2}  \exp(-\sum_{\ell=1}^d \xi_\ell^2).
$
We use tensor-product Gauss-Hermite quadrature rule to approximate Eq.~(\ref{expec}).
Denoting by $\{w_q\}_{q=1}^Q$ and $\{a_q\}_{q=1}^Q$ the weights and abscissae of the $Q$-point tensor-product Gauss-Hermite rule,  the approximation, denoted by $\widehat{\mathbb{E}}[\cdot]$ is given by
%%
%\[
%\widehat{\mathbb{E}}_{t_n}^{x}[g(X^{n+1})] := \sum_{\jj = 1}^J w_{\jj} \; g\Big(x+ b(t_n, x ) \Delta t + \sigma(t_n, x ) \sqrt{2\Delta t}\,a_{\jj} \Big),
%\]
%
\begin{equation}\label{quad}
\widehat{\mathbb{E}}\big[u^{n+1,J}(X_{n+1}^{t_n,x_j})\big] = \sum_{q = 1}^Q w_{q} \; u^{n+1,J}(z_{jq}^n)
\end{equation}
where 
%$z_{jq}^n$ is given by
\begin{equation}\label{e31}
z_{jq}^n := x_j+ b(t_n, x_j) \Delta t + \sigma(t_n, x_j) \sqrt{2\Delta t}\,a_{q}.
\end{equation}
Substituting Eq.~\eqref{quad} into Eq.~\eqref{e30}, completes our  numerical scheme for approximating the nodal value 
\begin{equation}\label{e304}
 u^n_j = \widehat{\mathbb{E}}\big[ u^{n+1,J}\big(X_{n+1}^{t_n,x_j}\big)\big],\;\; \text{ for } x_j \in \mathcal{S}\cap\mathcal{D},
\end{equation}
from where $u^{n,J}(x)$ can be obtained by interpolating $u^n_j$ using Eq.~\eqref{Lag11}.

\begin{rem}
%[The number of quadrature points $Q$]
\label{rem2}
Let $Q^* = Q^{1/d}$ denote the number of quadrature points in each dimension. If $u^{n+1,J}(\cdot)$ is sufficiently smooth, i.e., $\partial^{2J^*} u^{n+1,J}/\partial \xi_\ell^{2Q^*}$ is bounded for $\ell = 1, \ldots, d$, then the quadrature error is bounded by \cite{2013JSV...332.4403B}
\[
\left|\widehat{\mathbb{E}}[u^{n+1,J}(X_{n+1}^{t_n,x_j})] - {\mathbb{E}}[u^{n+1,J}(X_{n+1}^{t_n,x_j})]\right| \le C\frac{Q^*!}{2^{Q^*}(2Q^{*})!} (\Delta t)^{Q^*},
\]  
where the constant $C$ is independent of $Q^*$ and $\Delta t$. Note that the factor $(\Delta t)^{Q^*}$ comes from the $2{Q^*}$-th order differentiation 
of the function $u^{n+1,J}$ with respect to $\xi_\ell$ for $\ell = 1, \ldots, d$. Therefore, to achieve $\mathcal{O}(\Delta t)$ convergence only $Q^* = 3$ quadrature points in each dimension are needed.
\end{rem}

\begin{rem}\label{rem1}
Since only a finite number of quadrature points are needed to approximate $\mathbb{E}[\cdot]$, Assumption \ref{as1} is satisfied by generating a spatial mesh $\mathcal{S}$ such that all the quadrature points $z_{jq}^n$ defined in Eq.~\eqref{e31} for $x_j \in \mathcal{S}\cap \mathcal{D}$, $q=1, \ldots, Q$ are located in $\overline{\mathcal{D}}$. 
% Specifically, the spatial mesh $\mathcal{S}$ satisfies
%
% In practice, we define $\Delta x := \max_{i\in I_{a}}\Delta x_i$, the set $\{x_i | i\in I_{a}\}$  contains all grid points adjacent to boundary $\partial\mathcal{D}$, then $\Delta x$ is defined by, based on the quadrature rule in Eq.~\eqref{quad},
% %
% \begin{equation}\label{dx}
% \Delta x :=  \overline{b}\Delta t + \overline{\sigma}\sqrt{2\Delta t} \,\max_{j = 1, \ldots, J}|a_j|,
% \end{equation} 
% where $\overline{b}$ and $\overline{\sigma}$ are the upper bounds of $b$ and $\sigma$, respectively, and $\max_{j = 1, \ldots, J}|a_j|$ is a maximum value of the quadrature abscissae. 
% In this way, all the quadrature points
% $\{q_{ij}\}_{j=1}^J$
% used in Eq.~\eqref{quad} will locate in the domain $\overline{\mathcal{D}}$. 
\end{rem}

Summarizing, the proposed fully-discrete probabilistic scheme for the backward adjoint equation in Eq.~\eqref{e3} for a fixed terminal time $T_N \in \mathcal{T}$ consists of the following steps:
\begin{scheme}[The backward scheme for the adjoint equation]\label{s4:full}
Given the temporal spatial partition $\mathcal{T}_N\times \mathcal{S}$, the terminal condition $u^N(x_j)$ for $x_j \in \mathcal{S}$, and the boundary condition $u^n(x_j)$ for $x_j \in \mathcal{S}\cap\partial {\mathcal{D}}$. For $n = N-1, \ldots, 0$, the approximation of $u(t_n, x)$ is constructed via the following steps:
\begin{itemize}\itemsep0.0cm
\item Step 1: generate quadrature abscissae $\{z_{jq}^n\}_{q=1}^Q$ via Eq.~\eqref{e31}, for $x_j \in \mathcal{S} \cap \mathcal{D}$;
%\[
% \left\{q_{ij} := x_i+ b(t_n, x_i) \Delta t + \sigma(t_n, x_i ) \sqrt{2\Delta t}\,a_{j} \right\};
% \]
\item Step 2: interpolate $u^{n+1,J}(x)$ at the quadrature abscissae to obtain $\{u^{n+1,J}(z_{jq}^n)\}_{q=1}^Q$;
%\[
% \left\{v^{n+1,p}(x_i+ b(t_n, x_i) \Delta t + \sigma(t_n, x_i ) \sqrt{2\Delta t}\,a_{j})\right\};
%\]
%\item Step 2: compute the expectation $\widehat{\mathbb{E}}\left[v^{n+1}(X_{n+1}^{t_n,x_i})\right]$ by substituting $(q_{i,\jj}, v^{n+1,p}(q_{i,\jj}))$ into Eq.~\eqref{quad}
%
\item Step 3: compute the nodal values ${u_j^n}_{j=1}^J$ using the quadrature rule in Eq.~\eqref{e304};
\item Step 4: construct the interpolant $u^{n,J}(x)$ by substituting $\{u^n_j\}_{j=1}^J$ into Eq.~\eqref{Lag11}.
\end{itemize}
\end{scheme}

There is no stability condition, e.g., the CFL condition for the explicit finite difference method, forced on Scheme 1. Given a fixed spatial mesh $\mathcal{S}$, our method is stable with any $\Delta t$. As Fig.~\ref{ex3:f2} shown, when Assumption 1 failed, the error does not decay in the order of $\mathcal{O}(\Delta t)$, but it does not blow up neither. Moreover, the error of Scheme 1: $e_n = |u(t_n,x) - u^{n,J}(x)|$ is controlled by 
\begin{equation}\label{e_err}
    e_n \sim \mathcal{O}\left(\Delta t + \frac{(\Delta x)^{p+1}}{\Delta t}\right),
\end{equation}
where $\Delta x := \max_{i\in \mathcal{S} \cap \mathcal{D}}\Delta x_i$. The proof of the error estimate is out of the scope of this paper, the reader refer to works \cite{zhao2014new,Zhao:2009cr} for details. In addition, since all grid points $x_i$ are sufficiently far from the boundary $\partial \mathcal{D}$, i.e., $\Delta x$ is on the order of $(\Delta t)^{\frac{1}{2}}$, the use of PICHP interpolation method $(p=3)$ can provide sufficient accuracy near the boundary. Hence, our method achieves first order convergence in $\Delta t$ and second order convergence in $\Delta x$.

\subsection{Extension of Scheme 1 to computing the exit probability in $[0, T_{\max}]$}
%\subsection{An efficient algorithm for approximating the exit probability}
\label{sec:PT2}
Scheme \ref{s4:full} provides an efficient and accurate method to solve the adjoint equation in Eq.~\eqref{e3} for a fixed $T_N \in \mathcal{T}$, which
gives the approximation of the exit probability $P(T_N,x) = u(0,x)$ at one time instant. A naive strategy to obtain the full dynamics 
of $P(t,x)$ for $(t, x) \in \mathcal{T} \times\mathcal{S}$ would be to repeatedly solve the adjoint equation in Eq.~\eqref{e3} $M$ times. Although straightforward, this approach is  extremely inefficient. To address this issue, here we propose a novel strategy based on the explicit nature of the proposed probabilistic scheme. 

As a first step, we rewrite the interpolation (Step 2 of Scheme \ref{s4:full}) in  matrix form 
\begin{equation}\label{e34}
\mathbf u^{n+1}_{\rm quad} = \bm \Psi^n\,
\begin{pmatrix}
 \mathbf u^{n+1}\\
\mathbf{d}\mathbf u^{n+1}
\end{pmatrix}:= \bm \Psi^n \mathcal{F}^n  \mathbf u^{n+1},
\end{equation}
where $\mathcal{F}^n$ is the operation to obtain the vector $[\mathbf u^{n+1};\mathbf{d}\mathbf u^{n+1}]$. And $\mathbf{d}\mathbf u^{n+1}$ is a $Jn \times 1$ matrix that contains all partial derivatives at all grid points, where $n$ is the dimension. $\mathbf{u}^{n+1}$, $\mathbf{u}_{\rm quad}^{n+1}$ and $\mathbf \Psi^{n} $ are $J\times 1$, $JQ \times 1$ and $JQ \times (n+1)J$ matrices, respectively, i.e., 
\[
\mathbf u^{n+1} := 
\begin{pmatrix}
u_1^{n+1} \\
u_2^{n+1} \\
\vdots\\
u_J^{n+1}
\end{pmatrix}, 
\qquad
\mathbf u_{\rm quad}^{n+1} := 
\begin{pmatrix}
u^{n+1,J}(z_{11}^n) \\
u^{n+1,J}(z_{12}^n) \\
\vdots\\
u^{n+1,J}(z_{jq}^n) 
\end{pmatrix}, 
\]
\[
\mathbf \Psi^{n} := 
\begin{pmatrix}
\psi_1(z_{11}^n) & \cdots & \psi_J(z_{11}^n) & \phi_{11}(z_{11}^n) & \cdots & \phi_{1J}(z_{11}^n) &\cdots &  \phi_{n1}(z_{11}^n) & \cdots & \phi_{nJ}(z_{11}^n)\\
\psi_1(z_{12}^n) & \cdots & \psi_J(z_{12}^n) & \phi_{11}(z_{12}^n) & \cdots & \phi_{1J}(z_{12}^n) &\cdots & \phi_{n1}(z_{11}^n) & \cdots & \phi_{nJ}(z_{11}^n)\\
\vdots & &  \vdots & \vdots &\vdots & \vdots &  & \vdots & &\vdots\\
\undermat{\mbox{basis for } u^{n+1}}{\psi_1(z_{MJ}^n) & \cdots & \psi_J(z_{jq}^n)} & \undermat{\mbox{basis for } du^{n+1}/dx_1}{\phi_{11}(z_{MJ}^n) & \cdots & \phi_{1J}(z_{jq}^n)} & \cdots& \undermat{\mbox{basis for }du^{n+1}/dx_n}{\phi_{n1}(z_{11}^n) & \cdots & \phi_{nJ}(z_{11}^n)}
\end{pmatrix},
\]

\vspace{0.9cm}
\noindent and $\{z_{jq}^n\}$ is defined in Eq.~\eqref{e31}.
% \newpage

\vspace{0.2cm}
Similarly, we rewrite the quadrature (Step 3 in Scheme \ref{s4:full}) in matrix form
\begin{equation}\label{e35}
\mathbf u^{n} = \mathbf Q\, \mathbf u^{n+1}_{\rm quad},
\end{equation}
where $\mathbf{u}^{n}$ and $\mathbf Q$ are $J\times 1$ and $J \times JQ$ matrices, respectively, i.e.,
\begin{equation}\label{e38}
\mathbf n^{n} := 
\begin{pmatrix}
u_1^{n} \\
u_2^{n} \\
\vdots\\
u_J^{n}
\end{pmatrix}, 
\;\;\;
\mathbf Q := 
\begin{pmatrix}
w_1 & \cdots & w_Q\\
 & & & w_1 & \cdots & w_Q\\
 & & & & & &   \ddots \\
  & & & & & & &  w_1 & \cdots & w_Q\\
\end{pmatrix}.
\end{equation}
In this notation, the propagation from $\mathbf u^{n+1}$ to $\mathbf{u}^{n}$ is represented by
\begin{equation}\label{e36}
\mathbf u^n = \mathbf Q \mathbf \Psi^n\, \mathcal{F}^n  \mathbf u^{n+1}.
\end{equation}
By defining the terminal condition as 

\begin{equation}
\bm{\chi}(x) := \left\{
\begin{aligned}
0, \qquad &  x \in \mathcal{D},\\
1,\qquad &  x \in \mathcal{\partial D},\\
\end{aligned}
\right.
\end{equation}

the exit probability $P(t,x)$ for a fixed $t_n \in \mathcal{T}$ can be approximated by 
%
% \begin{equation}\label{e37}
% \mathbf p^n = \left(\prod_{\ell=0}^{n-1} {\mathbf Q \mathbf \Psi^\ell}\right) \bm{\chi},
% \end{equation}
\begin{equation}\label{e37}
\mathbf p^n = {\mathbf Q \mathbf \Psi^0 \mathcal{F}^0\left(\mathbf Q \mathbf \Psi^1 \mathcal{F}^1 \cdots \left( \mathbf Q \mathbf \Psi^{n-1} \mathcal{F}^{n-1} \bm{\chi}
\right)\right)} ,
\end{equation}
where $\mathbf{p}^n := (p^n_1, \ldots, p^n_J)^{\top} \approx (P(t_n, x_1), \ldots, P(t_n, x_J))$ and $\bm{\chi} := (\chi(x_1), \ldots, \chi(x_J))$. As aforementioned, the naive way to compute $\mathbf p^n$ is to run the recursive form in Eq.~\eqref{e37} $M$ times for $n = 0, \ldots, M-1$, which requires either storing all the $\mathbf \Psi^n$ or re-assembling $\mathbf \Psi^n$ for $n$ times. Since the assembling of $\mathbf \Psi^n$ requires interpolation on the spatial mesh $\mathcal{S}$, such strategy becomes very inefficient as the degrees of freedom $M$ increases. To address this issue, we developed the following algorithm that only assembles $\mathbf \Psi^n$ once as well as does not need to store the generated $\mathbf \Psi^n$ for the following time steps. 

% \newpage
% \noindent\makebox[\linewidth]{\rule{\textwidth}{0.5pt}}\\
% \vspace{-0.3cm}
% \newline {\bf Algorithm 1}
% \vspace{0.1cm}
% \newline1:\, {\bf Initialization}: $\mathbf G = {\rm zeros}(M, N+1)$, $\bm \chi = (\chi(x_1), \ldots, \chi(x_M))^{\top}$, $\mathbf P = {\rm zeros}(M, N)$;
% \vspace{0.1cm}
% \newline2: \,Generate the matrix $\mathbf Q$ in Eq.~\eqref{e38};
% \vspace{0.1cm}
% \newline3: \,Evaluate $\mathbf{G}(1:M, 1:N+1) = \bm \chi$;
% \vspace{0.1cm}
% \newline4: \,{\bf For} $n = N-1, \ldots, 0$
% \vspace{0.1cm}
% \newline5: \qquad Generate the interpolation matrix $\mathbf \Psi^n$ in Eq.~\eqref{e34};
% \vspace{0.1cm}
% \newline6: \qquad Update $\mathbf{G}(1:M, 1:n+1) = \mathbf{Q} \mathbf{\Psi}^n\, \mathbf{G}(1:M, 1:n+1) $;
% \vspace{0.1cm}
% \newline7: \qquad $\mathbf P(1:M, n+1) = \mathbf{G}(1:M, n+1)$;
% \vspace{0.1cm}
% \newline8: \,{\bf End} 
% \vspace{0.1cm}
% \newline9: \,{\bf Output}: $\mathbf{P}(1:M,1:N)$;\\
% \noindent\makebox[\linewidth]{\rule{\textwidth}{0.5pt}}
% %
%  \\

\noindent\makebox[\linewidth]{\rule{\textwidth}{0.5pt}}\\
\vspace{-0.3cm}
\newline {\bf Algorithm 1}
\vspace{0.1cm}
\newline1:\, {\bf Initialization}: $\mathbf P = {\rm zeros}(J, M)$, $\bm \chi = (\chi(x_1), \ldots, \chi(x_J))^{\top}$;
\vspace{0.1cm}
\newline2: \,Generate the matrix $\mathbf Q$ in Eq.~\eqref{e38};
\vspace{0.1cm}
\newline3: \,Evaluate $\mathbf{P}(1:J, 1:M) = {\rm repmat}(\bm \chi, 1, M)$;
\vspace{0.1cm}
\newline4: \,{\bf For} $n = M, \ldots, 1$
\vspace{0.1cm}
\newline5: \qquad Generate the interpolation matrix $\mathbf \Psi^{n-1}$ in Eq.~\eqref{e34};
\vspace{0.1cm}
\newline6: \qquad Generate the value matrix $\mathcal{F}^{n-1} \mathbf{P}(1:J, n:M)$ in Eq.~\eqref{e34};
\vspace{0.1cm}
\newline7: \qquad Update $\mathbf{P}(1:J, n:M) = \mathbf{Q} \mathbf{\Psi}^{n-1}\, \mathcal{F}^{n-1} \mathbf{P}(1:J, n:M)$;
% \vspace{0.1cm}
% \newline7: \qquad $\mathbf P(1:J, n+1) = \mathbf{G}(1:J, n+1)$;
\vspace{0.1cm}
\newline8: \,{\bf End} 
\vspace{0.1cm}
\newline9: \,{\bf Output}: $\mathbf{P}(1:J,1:M)$;\\
\noindent\makebox[\linewidth]{\rule{\textwidth}{0.5pt}}
 \\
 \vspace{-0.3cm}

% \noindent\makebox[\linewidth]{\rule{\textwidth}{0.5pt}}
% \vspace{-0.3cm}
% \newline {\bf Algorithm}
% \vspace{0.1cm}
% \newline1:\, {\bf Initialization}: $\mathbf P = {\rm zeros}(M, N+1)$, $ \chi = (\chi(x_1), \ldots, \chi(x_M))^{\top}$;
% \vspace{0.1cm}
% \newline2: \,Generate the matrix $\mathbf Q$ 
% \vspace{0.1cm}
% \newline3: \,Evaluate $\mathbf{P}(1:M, 1:N+1) =  \chi$;
% \vspace{0.1cm}
% \newline4: \,{\bf For} $n = N-1, \ldots, 0$
% \vspace{0.1cm}
% \newline5: \qquad Generate the interpolation matrix $\Psi^{n}$
% \vspace{0.1cm}
% \newline6: \qquad Update $\mathbf{P}(1:M, n+1:N) = \mathbf{Q} {\Psi}^{n}\, \mathbf{P}(1:M, n+1:N) $;
% \vspace{0.1cm}
% \newline7: \,{\bf End} 
% \vspace{0.1cm}
% \newline8: \,{\bf Output}: $\mathbf{P}(1:M,1:N+1)$;\\
% \noindent\makebox[\linewidth]{\rule{\textwidth}{0.5pt}}

The $n$-th column of the output $\mathbf P$ is the approximation of the exit probability $P(t,x)$ at the time step $t_{n} \in \mathcal{T}$ on the spatial mesh $\mathcal{S}$. 

% By using the intermediate matrix $\mathbf G$, we can carry the propagation information from $\mathbf Q \mathbf \Psi^n$ to the time steps $t_{n-1}, t_{n-2}, \ldots, 0$, such that we do not need to re-assemble or store $\mathbf \Psi^n$.

\subsubsection{GPU acceleration}
The main computational burden in Algorithm 1 is the matrix-matrix production for updating the matrix $\mathbf P$. At each time step, we can directly assemble the product $\mathbf Q \mathbf \Psi^n$.
% \begin{equation}\label{e40}
% \mathbf Q \mathbf \Psi^{n} := 
% \begin{pmatrix}
% \sum_{j=1}^J w_j \psi_1(q_{1j}) & \cdots & \sum_{j=1}^J w_j \psi_M(q_{1j}) \\
% \sum_{j=1}^J w_j \psi_1(q_{2j}) & \cdots & \sum_{j=1}^J w_j \psi_M(q_{2j}) \\
% \vdots & & \vdots\\
% \sum_{j=1}^J w_j \psi_1(q_{Mj}) & \cdots & \sum_{j=1}^J w_j \psi_M(q_{Mj}) \\
% \end{pmatrix}.
% \end{equation}
Due to the local support of the basis functions $\psi_j(x)$, the product, it is easy to see that $\mathbf Q \mathbf \Psi^{n}$ is a sparse matrix. By construction, the functions $\psi_j(x)$ can be evaluated in $\mathcal{O}(1)$ complexity, hence forming $\mathbf Q \mathbf \Psi^n$ has the complexity $\mathcal{O}(J \times Z)$ where $Z$ is the number of non-zeros on each column. In contrast, the complexity of the  matrix-matrix multiplication is $\mathcal{O}(J^2 \times Z)$, which dominates the total cost. 

In this work, we utilized the GPU architecture to accelerate the computation for updating $\mathbf P$. The vendor provided GPU sparse linear algebra library (cuSparse) \cite{cusparse} provides two sparse matrix-matrix functions, where the sparse matrix has to be stored in row compressed format, the dense input matrix can be given in either row or column major format, and the result is always given in column major form. In addition, the sparse matrix has to be on the left. The formulation for updating $\mathbf P$ in Algorithm 1 is very well suited for the cuSparse library. We implemented the GPU-enabled probabilistic scheme by coupling our code with the GPU version of the function approximation library TASMANIAN \cite{stoyanov2015tasmanian}. TASMANIAN provides acceleration via a set of custom CUDA kernels, while the performance of each kernel falls behind the vendor optimized library, the memory footprint is significantly lower since the matrix returned by the kernel has the correct format. The custom CUDA kernels provide a good compromise between performance and memory footprint. 

\section{Numerical examples}\label{sec:ex}
In this section we present three numerical examples illustrating and testing the proposed numerical method. 
The first example considers the computation of the exit time probability in a standard 1D Brownian motion. Since in this case the analytical expression of the exit probability is known, this example is used to 
demonstrate the convergence of our approach and the efficiency of the GPU-accelerated algorithm.
 The second example 
considers the exit time in a 2D advection-diffusion problem with a time dependent incompressible fluid velocity field. 
This example is used to compare the stability, accuracy, and computational cost of our  approach with the implicit and explicit finite different solution of the adjoint Fokker-Planck partial differential equation. 
The third example presents a 3D  plasma physics motivated problem and illustrates the seamless incorporation of nonuniform spatial grids in our scheme.

\subsection{The 1D Brownian motion}\label{sec:BM}
In this case the dynamics is governed by the  one-dimensional stochastic differential equation
\[
X_t = X_0 + W_t \quad \text{ with }  X_0 \in [0,L] \subset \mathbb{R},  
\]
and $t \in [0, T_{\max}]$.
 According to  Eq.~\eqref{e5}, the exit time probability $P(T,x)$ at $T \le T_{\max}$ is given by 
 $P(T,x) = u(t=0,x)$ where $u(t,x)$ is the solution of the terminal value problem
\begin{equation}\label{e44}
\begin{aligned}
\frac{\partial u(t,x)}{\partial t} + \frac{1}{2} \frac{\partial^2 u(t,x)}{\partial x^2} & = 0 \quad \text{ for }\;\; x \in [0,L], t < T,\\
u(t,x) & = 1  \quad \text{ for }\;\; x =\{0,L\}, t<T,\\[4pt]
 u(T,x) & = 0  \quad \text{ for }\;\; x \in [0,L] .\\
\end{aligned}
\end{equation}
From the known exact analytical solution of Eq.~(\ref{e44}) it follows that 
\begin{equation}\label{e45}
P(T,x) = 1 - \sum_{n=1}^\infty \frac{4}{n\pi} \sin\left( \frac{n \pi x}{L}  \right) \exp\left[-\frac{1}{2} \left(\frac{n \pi}{L}\right)^2 T   \right].
\end{equation}

\begin{figure}[h!]
\begin{center}
\includegraphics[scale =0.3]{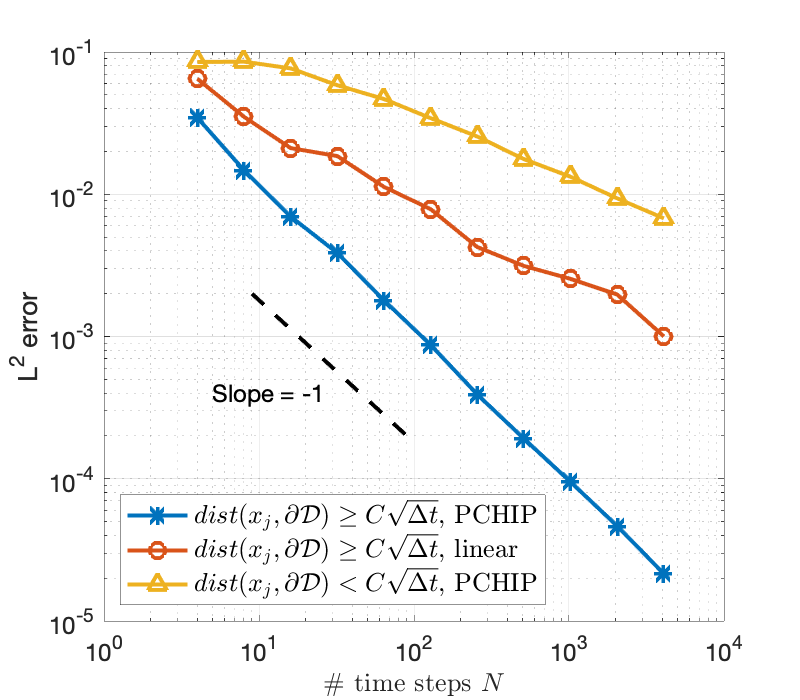}
\vspace{-0.2cm}
\caption{Error decay in the computation of $P(T_{\max},x)$, for $T_{\max} = 3$ and $L=10$. The proposed method achieves  $\mathcal{O}(\Delta t)$ convergence when using 3-point Gauss-Hermite rule, PCHIP interpolation and enforcing $dist(x_j, \partial \mathcal{D}) \ge C\sqrt{\Delta t}$ in Eq.~\eqref{e301} for $x_j \in \mathcal{S}\cap \mathcal{D}$. In comparison, the use of linear interpolation cannot provide sufficient accuracy near the boundary $\partial \mathcal{D}$; violating the condition in Eq.~\eqref{e301} makes the probability $\mathbb{P}\left(\tau_{t_n,x} \le t_{n+1}\right)$ is too big to be neglected for the grid points near $\partial \mathcal{D}$.}\label{ex3:f2}
\end{center}
\end{figure}

Figure \ref{ex3:f2} shows the first-order convergence of our method with respect to $\Delta t$ for the case
$L = 10$ and $T_{\max} = 3$, with the error given by the Frobenius norm of the difference of the numerically computed solution and the exact solution in Eq.~(\ref{e45}).  
To balance the errors from quadrature and interpolation, we use 3-point Gauss-Hermite rule and enforce the condition on the spatial mesh $\mathcal{S}$ which is discussed in Assumption \ref{as1} and Remark \ref{rem1}. In comparison, when replacing PCHIP interpolation with linear interpolation, the error decays slower, due to the enforcement of the condition $dist(x_j, \partial \mathcal{D}) \ge C\sqrt{\Delta t}$ in Eq.~\eqref{e301} near the boundary $\partial \mathcal{D}$. On the other hand, 
when the condition in Eq.~\eqref{e301} is violated, i.e. when $dist(x_j, \partial \mathcal{D}) < C\sqrt{\Delta t}$, 
the error also decays slower because the probability $\mathbb{P}\left(\tau_{t_n,x} \le t_{n+1}\right)$ is too big to be neglected for the grid points near $\partial \mathcal{D}$. 

\begin{figure}[h!]
\begin{center}
\includegraphics[scale =0.3]{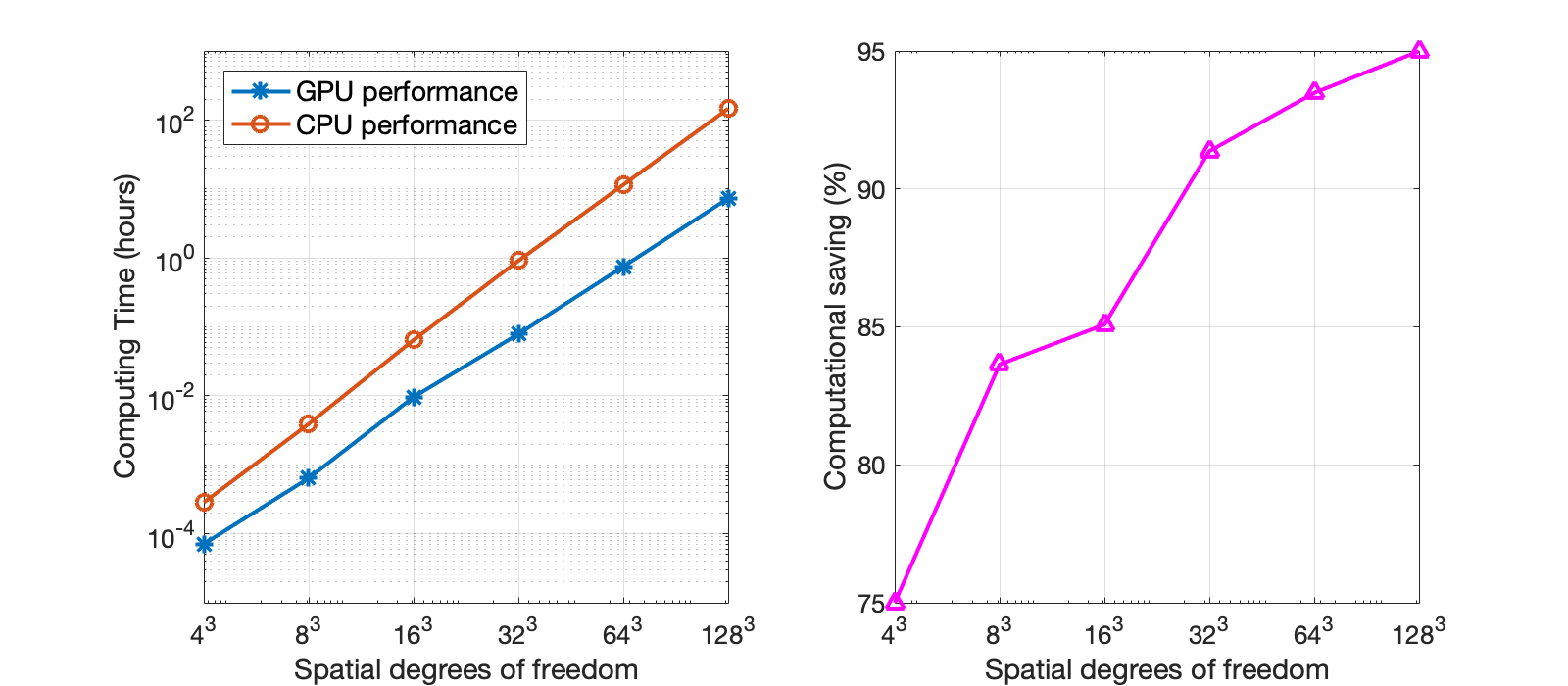}
\vspace{-0.3cm}
\caption{Performance comparison of GPU and CPU implementation of Algorithm 1 of the proposed method for the exit time probability in a $1D$ Brownian motion.}\label{ex3:f3}
\end{center}
\end{figure}

Figure \ref{ex3:f3} illustrates the GPU acceleration of Algorithm 1. For comparison we implemented both a CPU and GPU version in Matlab, and utilized a Matlab interface to call the GPU-based matrix-matrix multiplication module in TASMANIAN. The results were generated on a workstation with Intel Xeon W-2225 CPU and Nvidia Quadro P1000 GPU. The time reported in Figure \ref{ex3:f3} is the wall clock time for solving the entire problem. It is observed that the GPU-accelerated algorithm  reduces the total computing time by up to 95\%. 

\subsection{A 2D advection diffusion problem}\label{sec:fluid}
% in $\mathcal{D} :=  [-\pi,\pi]\times[0,L]$
%The purpose of this example is to compare our approach with finite difference scheme in stability, accuracy and computational cost. We consider the exit probability of a two-dimensional fluid problem that is described by the following advection-diffusion equation
In this example we consider the transport of a scalar field $u$ modeled by the advection-diffusion equation
\begin{equation}\label{ex2model}
\begin{aligned}
\frac{\partial u(t,{\bf x})}{\partial t} + P_e {\bf v}\cdot  \nabla u(t,{\bf x}) &= \frac{1}{2} \nabla^2 u(t,{\bf x}),
\end{aligned}
\end{equation}
for $t\in[0,T_{max}]$ in the 2D domain ${\bf x}\in \mathcal{D} :=  [-\pi,\pi]\times[0,L]$,  with time-dependent incompressible velocity field  ${\bf v}=v_{x_1} {\bf e}_{x_1}+v_{x_2} {\bf e}_{x_2}$, where
\begin{eqnarray}
v_{x_1} &= &- \pi \cos (\pi x_2)\left[ \sin (n x_1) + \epsilon f(t) \cos (n x_1) \right],  \\ 
v_{x_2} &=& n \sin (\pi x_2)\left[\cos (n x_1)- \epsilon f(t) \sin (n x_1) \right],\\
f(t)&=&\cos (\omega t). 
\end{eqnarray}
% The corresponding SDEs, which defined in $[0,T_{\max}]\times \mathcal{D}$, are
% \begin{eqnarray}
% d x &=& P_e\, v_x d t + d W_x \\
% d y &=& P_e\, v_y d t + d W_y \,.
% \end{eqnarray}
This cellular flow velocity field \cite{chandrasekharhydrodynamic} is a commonly used model to study  transport in Rayleigh-Bernard convection, see e.g. \cite{young1989anomalous}. 
The corresponding 2D, non-autonomous SDE system is
\begin{equation}\label{ex3}
\left\{
\begin{aligned}
dx_1 &=  P_e\, v_{x_1} (x_1,x_2,t) \, dt +\, dW_1,\\ 
dx_2 &= P_e\, v_{x_2} (x_1,x_2,t) \,  dt + \, dW_2 \,   ,
\end{aligned}
\right.
\end{equation}
where $dW_1$ and $dW_2$ are independent Wiener processes (Brownian motions).

We use dimensionless variables with ${\bf x}$  normalized using the length scale $L$, and $t$  normalized using the diffusion time scale $T=L^2/D$, where $D$ is the diffusivity. The P\'eclet number $P_e$ is defined as $P_e= VL/D$ where $V$ is the velocity scale. The free parameters of the model are $P_e$, $\epsilon$, $\omega$, and $n$.  The regimen $P_e \gg 1$ ($P_e \ll 1$) corresponds to an advection (diffusion) dominated regime. On the other hand, the limit $\omega \ll 1$ ($\omega \gg 1$) corresponds to a perturbation with a period significantly longer (shorter) than the diffusion time scale. The parameter $\epsilon$ determines the size of the perturbation and the number of cells in the flow pattern is $2 n$.  For the numerical simulations we take: $P_e=10$, $\epsilon=0.15$, $\omega=2 \pi/10$, $n=2$, $T_{max} = 0.05$, and $L = 1$. The boundarey conditions are 
periodic in $x_1$, and Dirichlet, $P(t,x_1,0) = P(t,x_1,1) = 1,$ in $x_2$.

\begin{figure}[h!]
\begin{center}
\includegraphics[scale =0.3]{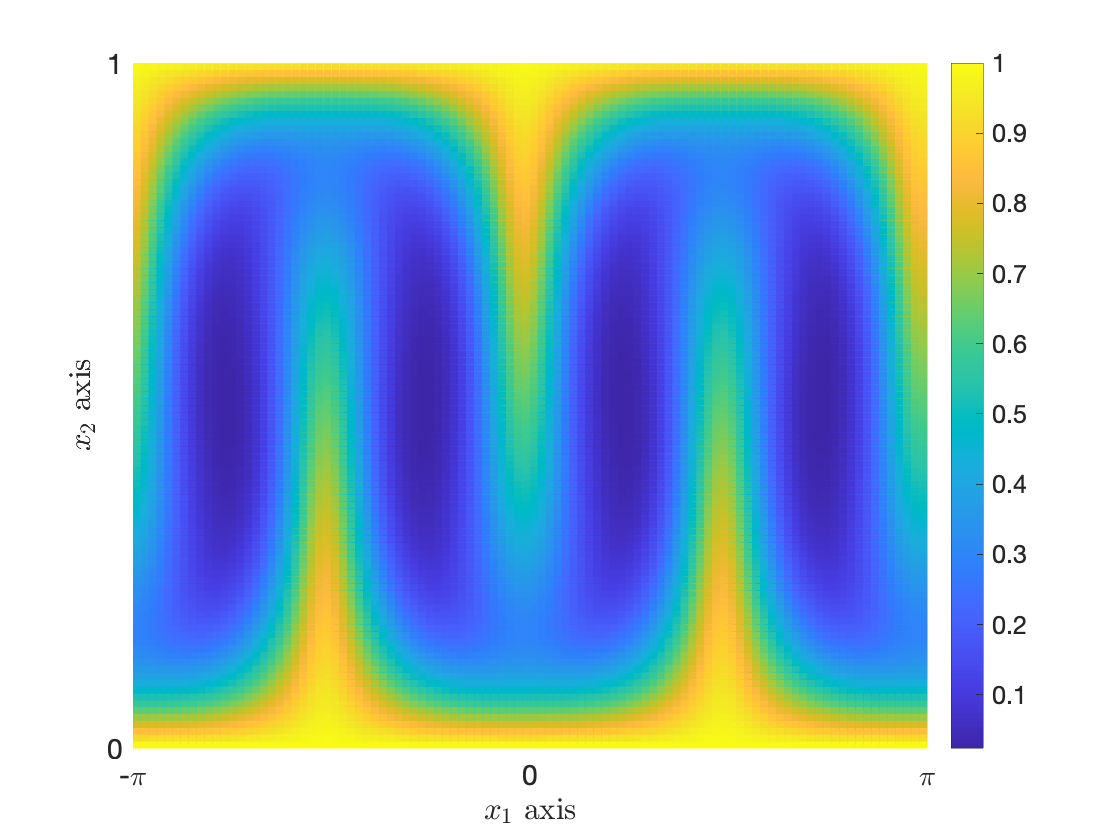}
\vspace{-0.3cm}
\caption{the exit probability $P(\bm x, T_{\max})$ for the advection-diffusion problem in Eq.~\eqref{ex2model}.}\label{ex2:f1}
\end{center}
\end{figure}

In this fluid mechanics problem the exit time probability corresponds to the probability that a Lagrangian element of the passive tracer $u$ located at ${\bf x}$ reaches the top ($x_2=1$) or the bottom ($x_2=0$) boundary. Since $n=2$, in this case the unperturbed, $\epsilon=0$,  flow exhibits four convection rolls with centers  at the stagnation elliptic points of the flow field located at $(x_1,x_2)=(\pm \pi/4,1/2)$ and
 $(x_1,x_2)=(\pm 3 \pi/4,1/2)$. The cells of the convection rolls are flanked by the vertical separatrices joining the hyperbolic stagnation points of the flow
located at $(x_1,x_2)=(\pm \pi/2,0)$, $(x_1,x_2)=(\pi,0)$ and
$(x_1,x_2)=(0,0)$ with those located at
$(x_1,x_2)=(\pm \pi/2,1)$, $(x_1,x_2)=(\pm \pi,1)$ and
$(x_1,x_2)=(0,1)$.
In the absence of a time dependent perturbation, $\epsilon=0$, the migration towards the $x_2=0$ and $x_2=1$ boundaries is due to diffusion. However, when $\epsilon \neq 0$ the flow field exhibits chaotic advection \cite{aref2002development,solomon1988chaotic} and transport  is enhanced specially near the separatrices of the flow. 
This phenomenology is  in agreement by the computed exit time probability shown in Fig.~\ref{ex2:f1}. In particular, the exit time probability is small in the vicinity of the elliptic points of the flow (i.e., the core of the convection rolls) and large near the separatrices and the hyperbolic stagnation points.

\begin{figure}[h!]
\minipage{0.33\textwidth}
  \includegraphics[width=\linewidth]{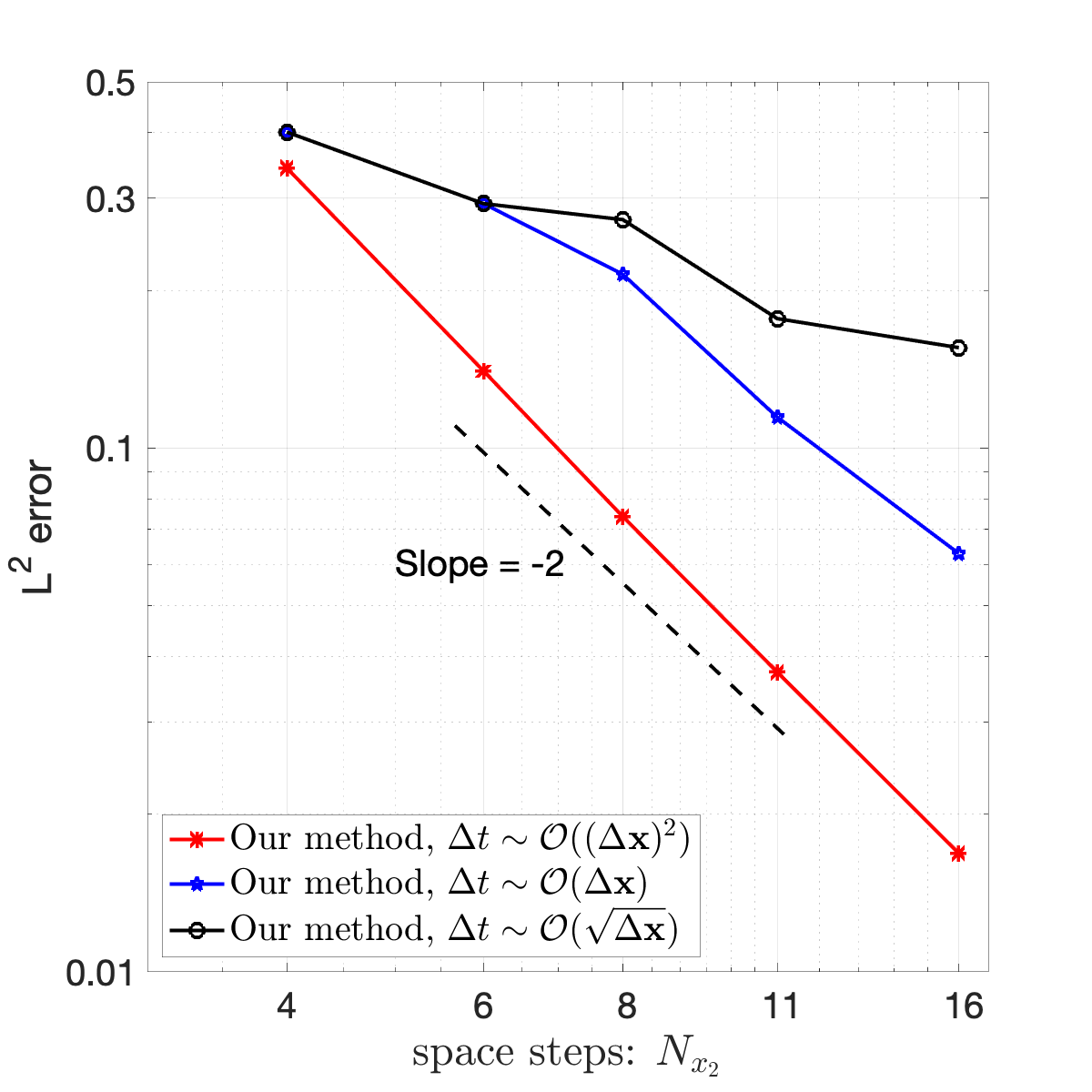}
\endminipage\hfill
\minipage{0.33\textwidth}
  \includegraphics[width=\linewidth]{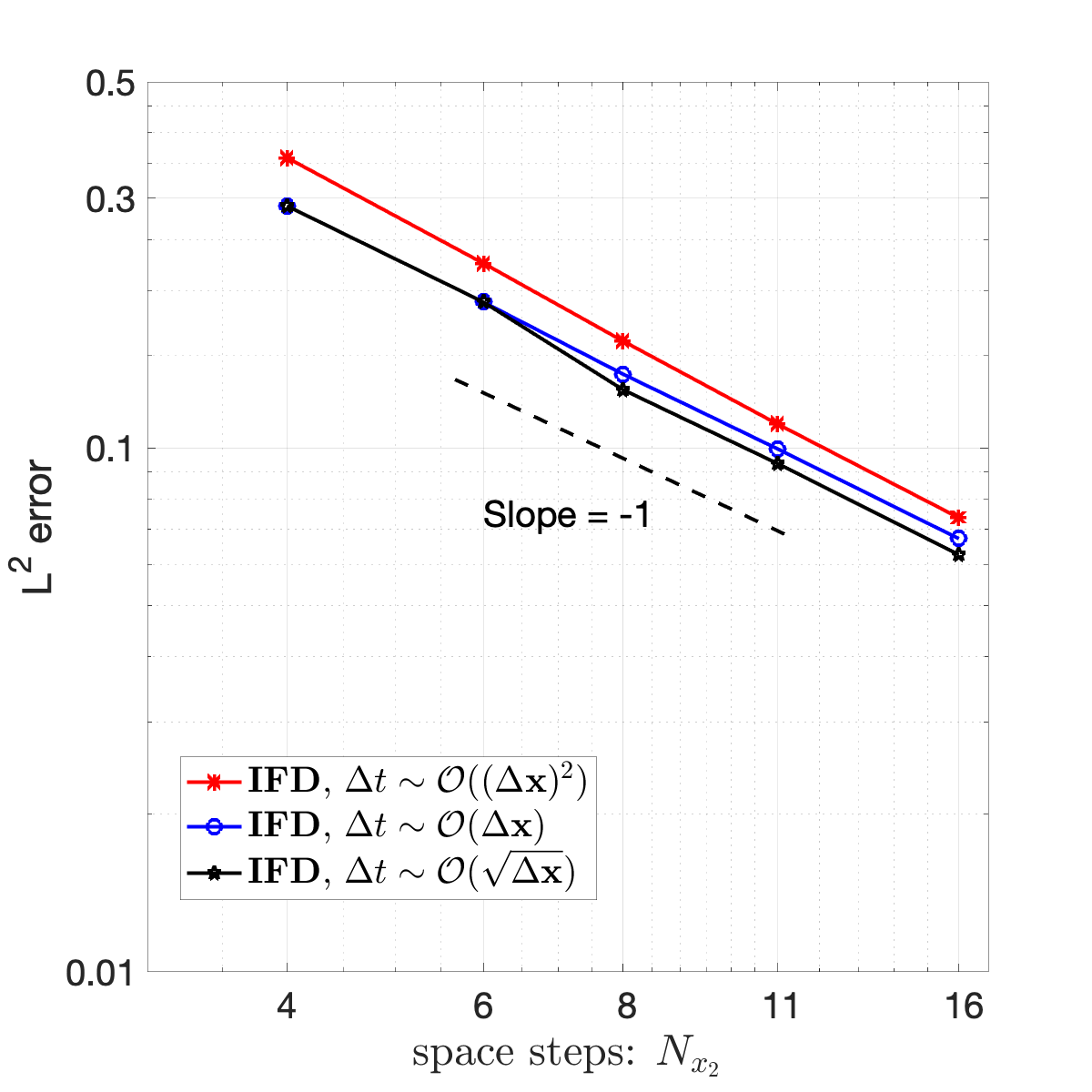}
\endminipage\hfill
\minipage{0.33\textwidth}%
  \includegraphics[width=\linewidth]{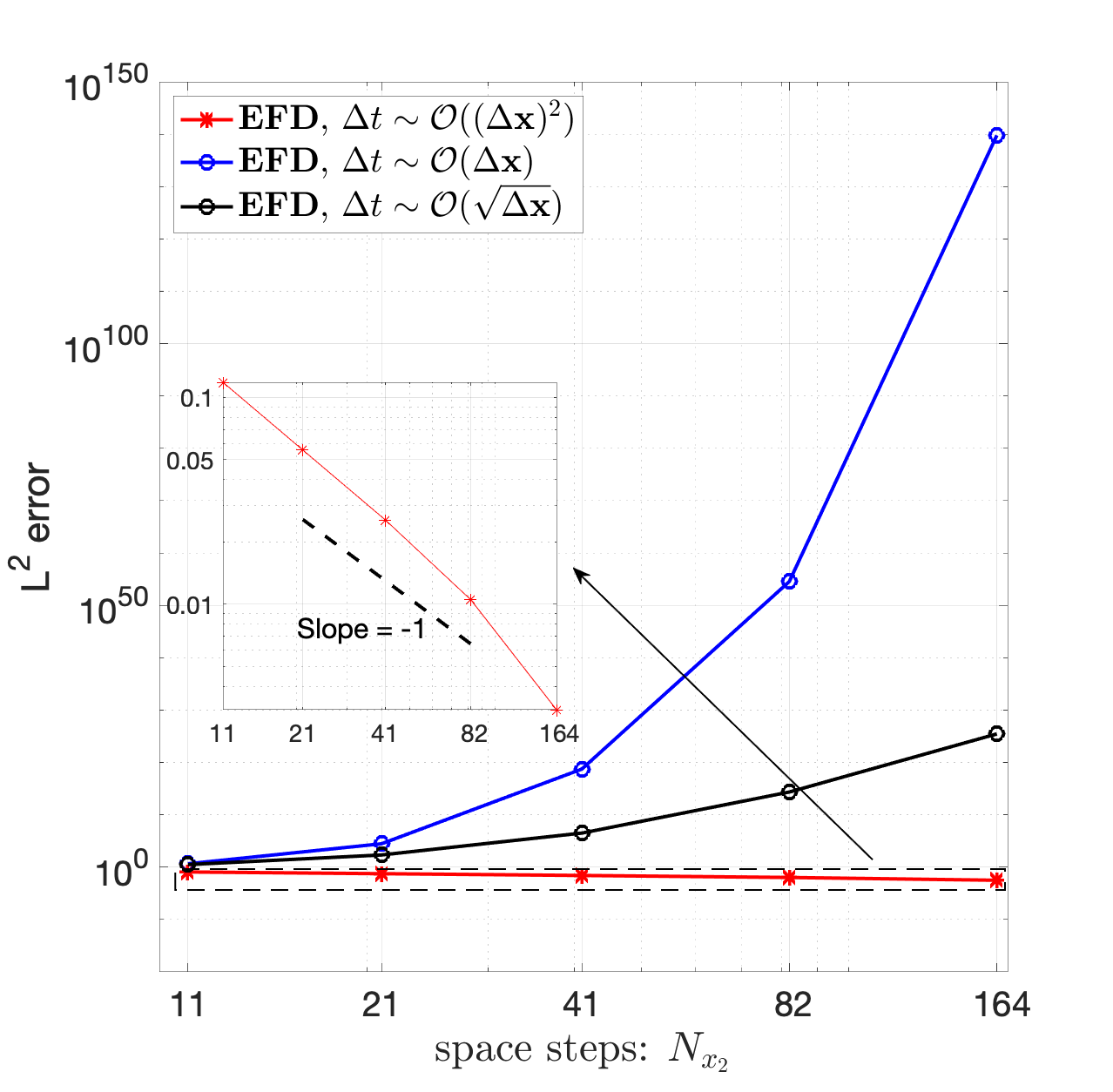}
\endminipage
\caption{The proposed method is unconditional stable and exhibits second-order convergence in $\Delta x$ when 
Assumption \ref{as1} is satisfied. In comparison, the implicit finite difference (IFD)
and the explicitly finite difference (EFD) schemes exhibit  first-order convergence. In addition, the EFD 
is constrained by the CFL condition to guarantee stability.}
\label{bmcefd}
\end{figure}

Figure \ref{bmcefd} presents a comparative study of the error decay as function of the grid resolution. The error was quantified using the Frobenius norm of the difference of the computed solution and a reference solution. Since there are not known analytical solutions for this problem, we used as reference a high resolution numerical integration of Eq.~(\ref{ex2model}) using an explicit finite difference (EFD) method. The reference solution was fully converged in the sense that it did not change when reducing the grid resolution and time step. 
As the left panel in 
Fig.~\ref{bmcefd} shows, our method achieves second-order convergence in $\Delta {\bf x}$, when Assumption \ref{as1} is satisfied. When Assumption \ref{as1} is not satisfied the error exhibits a slower decay but it does not blow up because the method is unconditionally stable. 
In contrast, as the middle panel shows, the implicit finite difference (IFD) method can only achieve 
first-order convergence in $\Delta x$ provide a first-order upwind scheme is used to guarantee stability
(the second-order central scheme is unstable). The EFD, shown in the right panel, can also exhibit first order convergence but it is significantly constrained by the CFL stability condition. 
The extension of first-order upwind schemes to higher order faces technical challenges at grid points adjacent to the boundary. For schemes using stencils involving values outside of the domain, there should be additional strategies to figure them out. In comparison, our method inherently simulates the direction of propagation of information, with the quadrature points incorporating the direction of advection.

\begin{figure}[h!]
\center
  \includegraphics[scale =0.25]{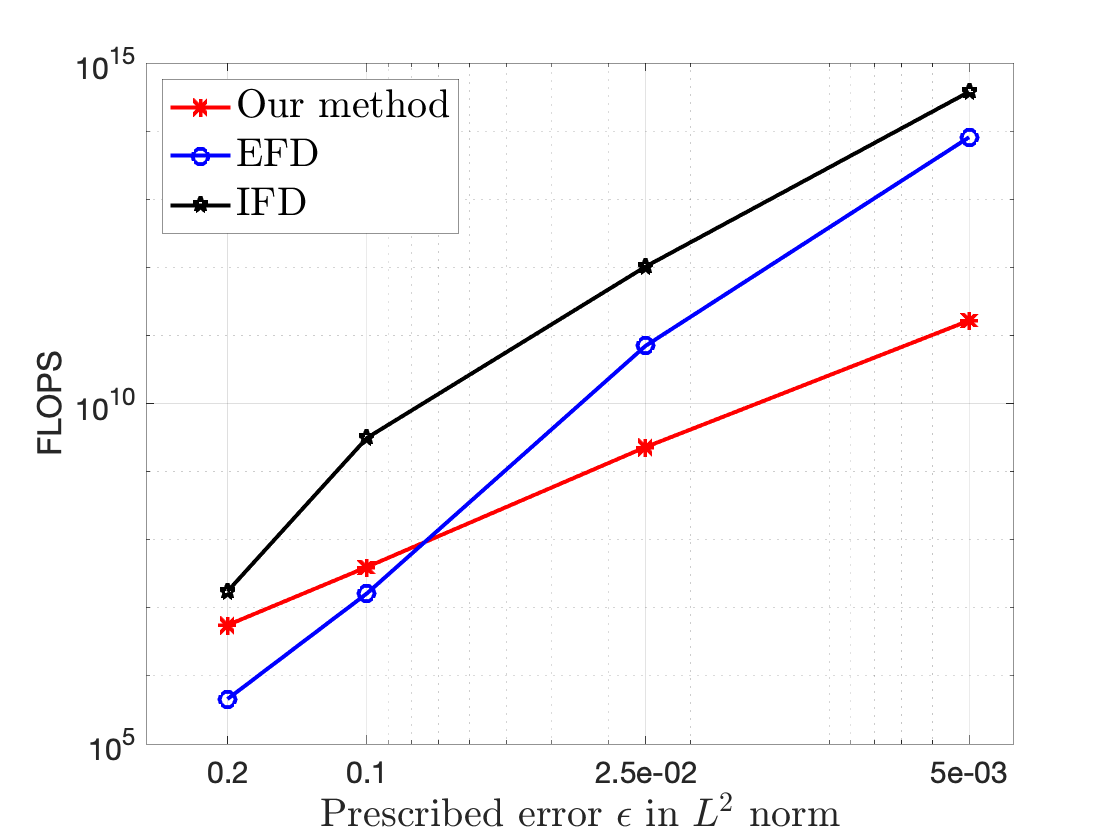}
  \caption{Comparison of the computational cost of the proposed method with  EFD and IFD methods as function of the prescribed error. To achieve comparable  small errors, EFD and IFD methods require several orders of magnitude more operations. This is because finite difference schemes require a double temporal iteration to recover the full dynamics of the exit probability. EFD are more efficient than IFD, but due to stability constrains, they require a large amount of time steps to achieve small errors.} 
  \label{Flops}
\end{figure}

Figure \ref{Flops} compares the computational cost of the proposed method with the EFD and IFD methods.
It is observed that, depending on the prescribed error, the new method requires up to several orders of magnitude less operations than finite difference methods for the computation of the full dynamics of the exit time probability in $t\in [0,T_{max}]$. This is because our method only requires a single temporal iteration for computing the exit probability for all time instants in the interval $[0,T_{max}]$. In contrast, the IFD scheme requires a double temporal iteration since it needs to solve the adjoint equation repeatedly to recover the full dynamics of the exit time probability. The EFD scheme is not much better due to the 
slow convergence rate, $\sqrt{\Delta t} \sim \Delta {\bf x} \sim \epsilon$, and the need to satisfy the CFL condition.
%, the EFD scheme needs much more time and space steps for a prescribed error than our method. For a prescribed error $\epsilon$, the cheapest flop is given by the corresponding temporal-spatial convergence rate. Results are shown in figure \ref{Flops}.

% Lastly, we move on to the comparison of the computational cost. The competitive advantage of the BMC is able to handle the full dynamics without solving the adjoint parabolic problem repeatedly compared to the implicit finite difference method. So the IFD is very expensive to obtain several exit probabilities compared with EFD and BMC. The main computational cost in the BMC algorithm is the propagation process, where we generate a matrix $\Psi^n$ which is on the order of $\mathcal{O}\left((N_x\cdot N_y)^2\right)$. We compare the least temporal-spatial steps $N_t, N_x, N_y$ for solving the exit probability at time $T$ for the prescribed errors $\epsilon$. As Figure \ref{Flopst},\ref{Flopsx} shown, for a prescribed errors $\epsilon$,  based on the globally convergence rate, we have 
% \begin{equation}
% \begin{aligned}
% &BMC: \Delta t \sim (\Delta x)^2 \sim (\Delta y)^2 \sim \epsilon\,;\\
% &\quad EFD: (\Delta t)^{1/2} \sim \Delta x \sim \Delta y \sim \epsilon\,;\\
% &\quad IFD: \Delta t\sim \Delta x \sim \Delta y \sim \epsilon\,.
% \end{aligned}
% \end{equation}

\subsection{A 3D runaway electron problem}\label{e200}

To demonstrate the seamless integration of the proposed method with nonuniform spatial grids, we consider a three-dimensional exit time probability problem motivated by plasma physics. 
The model of interest describes the acceleration of electrons in a magnetically confined  plasma in a toroidal chamber (tokamak) in the presence of an electric field taking into account collisional and radiation damping. The main physics problem is to understand the conditions under which the damping cannot overcome the electric field acceleration causing the electrons to ``runaway" achieving arbitrarily large energies. Understanding this problem is critical for the safe operation of nuclear fusion reactors, because if these so-called runaway electrons (RE) are not avoided or controlled they can damage the plasma facing components of the reactor. 
This is a very complex problem, and here, as a proof-of-principle calculation, we consider a 
simplified model describing the evolution of 
the  magnitude of the relativistic momentum of the electron, denoted by $p$, the cosine of the pitch angle,  $\xi=\cos \theta$, 
where $\theta$ is the angle between the momentum and the direction of the magnetic field used to confine the plasma, and the minor radius, $r$, of the toroidal confinement region. 
The time evolution of these variables is governed by the stochastic differential equations  
\begin{equation}\label{ex32d}
\left\{
\begin{aligned}
dp &=  \left [E \xi\, - \frac{\gamma p}{\tau}(1-\xi^2)-C_F(p) +\frac{1}{p^2}\frac{ \partial }{\partial p} \left(p^2 C_A(p)\right) \right ] dt + \sqrt{2 C_A(p)} \, dW_p,\\ 
d \xi &= \left[ \frac{E \left(1-\xi^2\right)}{p} - \frac{\xi (1-\xi^2)}{\tau \gamma}
-2 \xi \frac{C_B(p)}{p^2} \right ] dt + \frac{\sqrt{2 C_B(p)}}{p}  \,  \sqrt{1-\xi^2}\, dW_\xi, \,  \\
% \\ d \phi &=&  \frac{\sqrt{2 C_B}}{ p \sqrt{1-\xi^2}} \,  dW_\phi \,
dr & =  \left[V(r) + \frac{\partial D(r,p)}{\partial r}\right]dt + \sqrt{2D(r,p)} dW_r \, .
\end{aligned}
\right.
\end{equation}
where $dW_p$, $dW_\xi$ and $dW_r$ are independent Wiener processes (standard Brownian motions), $E$ is the electric field, and $\tau$ the radiation damping time scale. The collision operator is determined by the functions $C_A$, $C_B$, and $C_F$ defined as 
\begin{eqnarray}
C_A (p) &=& \bar{\nu}_{ee} \, \bar{v}_T^2 \,\,\frac{\psi(y)}{y},  \nonumber
 \\
C_B (p)&=& \frac{1}{2} \,\bar{\nu}_{ee} 
\, \bar{v}_T^2 \, \, \frac{1}{y}  \left[ Z + \phi(y)- \psi(y) +  \frac{y^2}{2} \delta^4 \right], \nonumber\\[0.3cm]
C_F (p)&=&2\,\bar{\nu}_{ee}  \, \bar{v}_T \, \psi(y), \,  \nonumber
\end{eqnarray}
$$
\phi(y)=\frac{2}{\sqrt{\pi}} \int_0^y e^{-s^2} ds \, ,\qquad
\psi(y)=\frac{1}{2 y^2} \left[ \phi(y)-y \frac{d \phi}{dy} \right],  \nonumber
%\qquad x=\frac{\tilde{v}_T}{v_T}\frac{p}{\gamma}\, ,
$$
$$
y = \frac{p}{\gamma},\qquad \gamma = \sqrt{1 + (\delta p)^2}, \qquad \delta = \frac{v_T}{c} = \sqrt{\frac{2T}{mc^2}} .
$$
with $Z$ and $c$ denoting the ion effective charge and the speed of light, respectively. 
Further details on the model can be found in Ref.~\cite{zhang2017backward} and references therein. 
The models for the velocity, $V$, and the diffusivity, $D$, in the equation governing the dynamics of $r$ are 
\begin{equation}
V(r) =   V_0\mathbbm{1}_{\{r>r_D\}},\quad D(r,p) = \frac{D_0}{2}[1+f(r)]e^{-(p/\Delta p)^2}, \quad f(r) = \tanh{\left(\frac{r-r_D}{W}\right)} \, .
\end{equation}
In the calculations presented we use the following typical model parameters
\begin{equation}
\begin{aligned}
\label{parameters}
%T_{\max} & = 120, \;\; p_{\min} = 2, \;\; p_{\max} = 50,\;\; Z = 1,\;\;  \tau = 10^5 \\
 & Z = 1,\;\;  \tau = 10^5 ,\;\;  \delta = 0.042,\;\;  E = 0.3, \;\; \bar{\nu}_{ee} = 1, \;\; \bar{v}_T =1 \\
 & r_D  =  0.7,\;\;   W = 10^{-2},\;\;  \Delta p = 20,\;\;  V_0 = 0.003 ,\;\;  D_0 = 10^{-4}
\end{aligned}
\end{equation}
The integration domain is $p\in (p_{\min}, p_{\max})$, $\theta \in (0, \pi)$, and $r\in (0,1)$. In this context, the probability of the exit time represents the probability that an electron located at $r$ with momentum $p$ and pitch angle $\theta$, ``runs away" and exits the integration domain by crossing the $p= p_{\max}$ boundary (entering a high energy region)  at a time less than or equal to $T_{\max}$. Based on this interpretation, in this section  we will refer to the  probability of the exit time as the 
``runaway probability", $P_{RE}$. For the calculations we  use 
$(p_{\min}, p_{\max})=(2, 50)$,  and $T_{\max}=120$.

\begin{figure}[h!]
\center
  \includegraphics[scale =0.4]{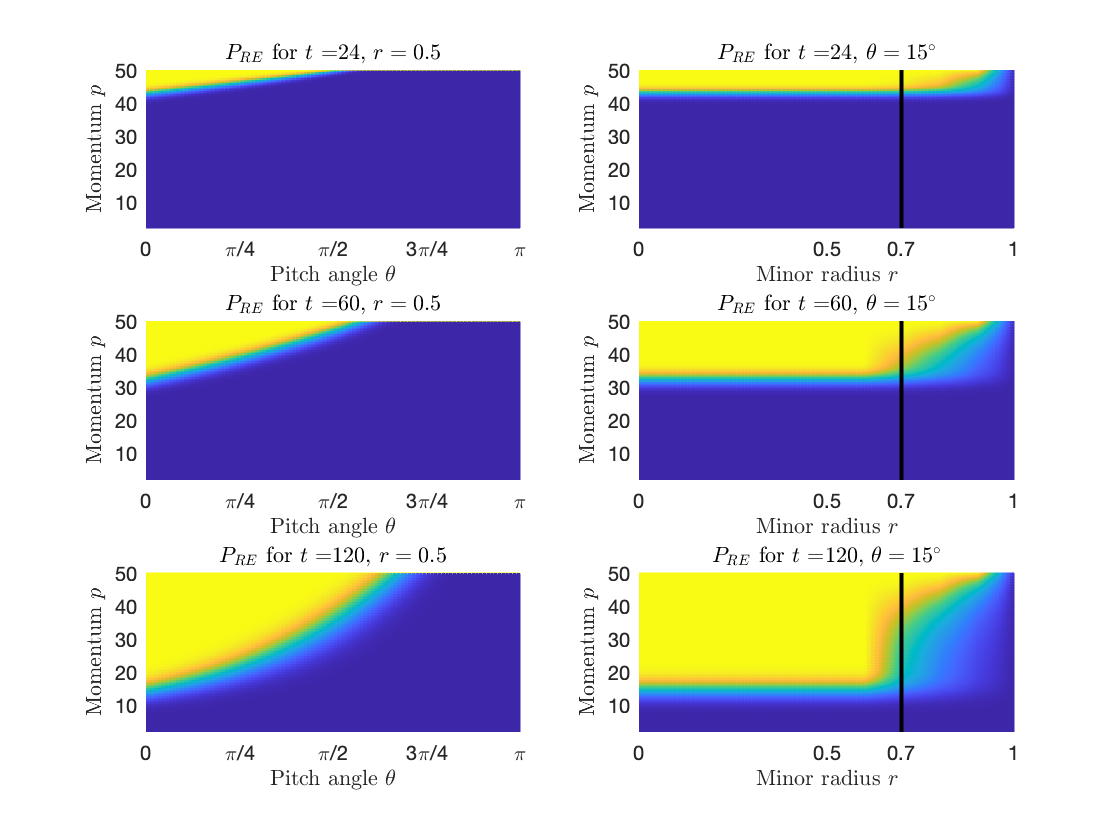}
  \caption{
Computation of runaway probability, $P_{RE}$, using a {\em low resolution uniform grid} ($\Delta r_s=0.1$).  
Left column shows $P_{RE}$ in the $(\theta,p)$ plane at $r=0.5$.
Right column shows $P_{RE}$ in the $(r,p)$ plane at $\theta=15^o$. 
The rows correspond to the three successive times $t=24$, $60$, and $120$. 
The use of a coarse grid leads to numerical diffusion causing an artificial spreading of the boundary separating the $P_{RE} \sim 0$ (blue)  and $P_{RE} \sim 1$ (yellow) boundary around $r=r_D=0.7$ (black vertical line).} 
  \label{uniform}
\end{figure}

\begin{figure}[h!]
\center
  \includegraphics[scale =0.4]{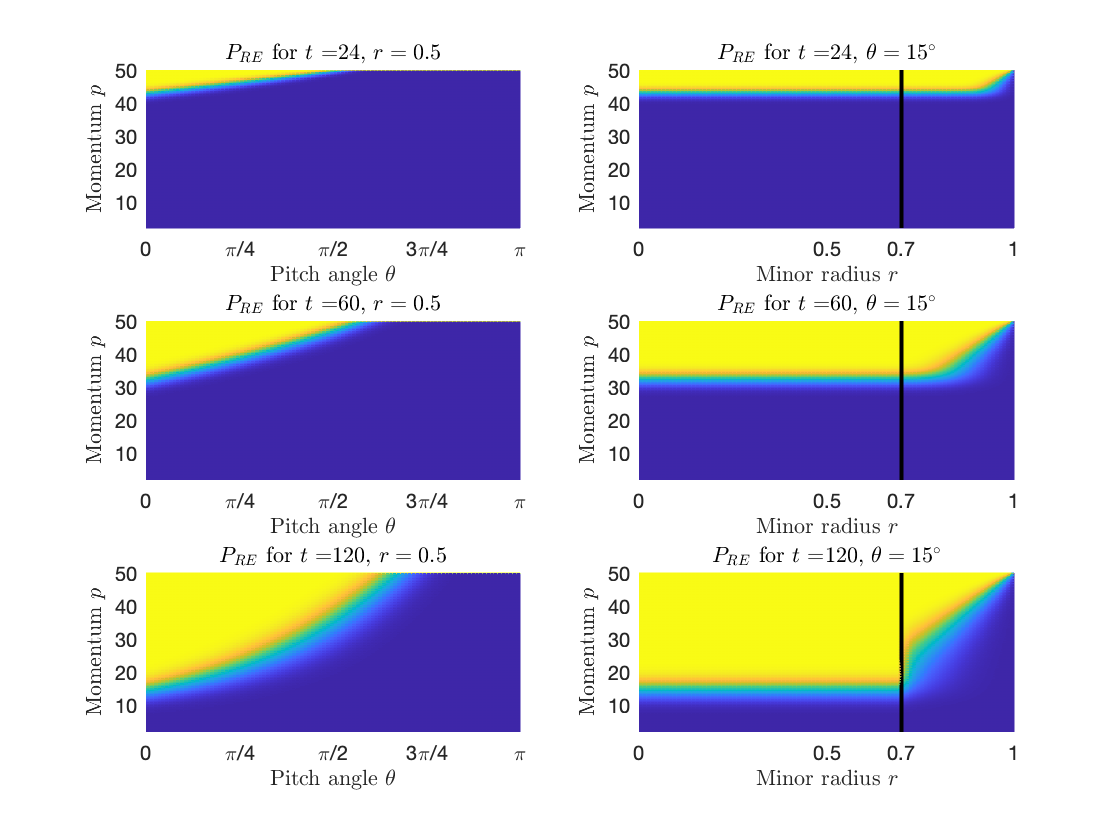}
  \caption{
 Same as Fig.~\ref{uniform} but using a {\em high resolution uniform grid} ($\Delta r_d = 0.0025$).
To eliminate the numerical diffusion shown Fig.~\ref{uniform} and recover the expected sharp transition between the $P_{RE} \sim 0$ (blue)  and $P_{RE} \sim 1$ (yellow) boundary around $r=r_D=0.67$ (black vertical line) using a uniform grid requires a reduction of the grid spacing by more than two orders of magnitude. This global uniform refinement leads to a significant increase of the computational cost} 
  \label{denseuni}
\end{figure}

\begin{figure}[h!]
\center
  \includegraphics[scale =0.4]{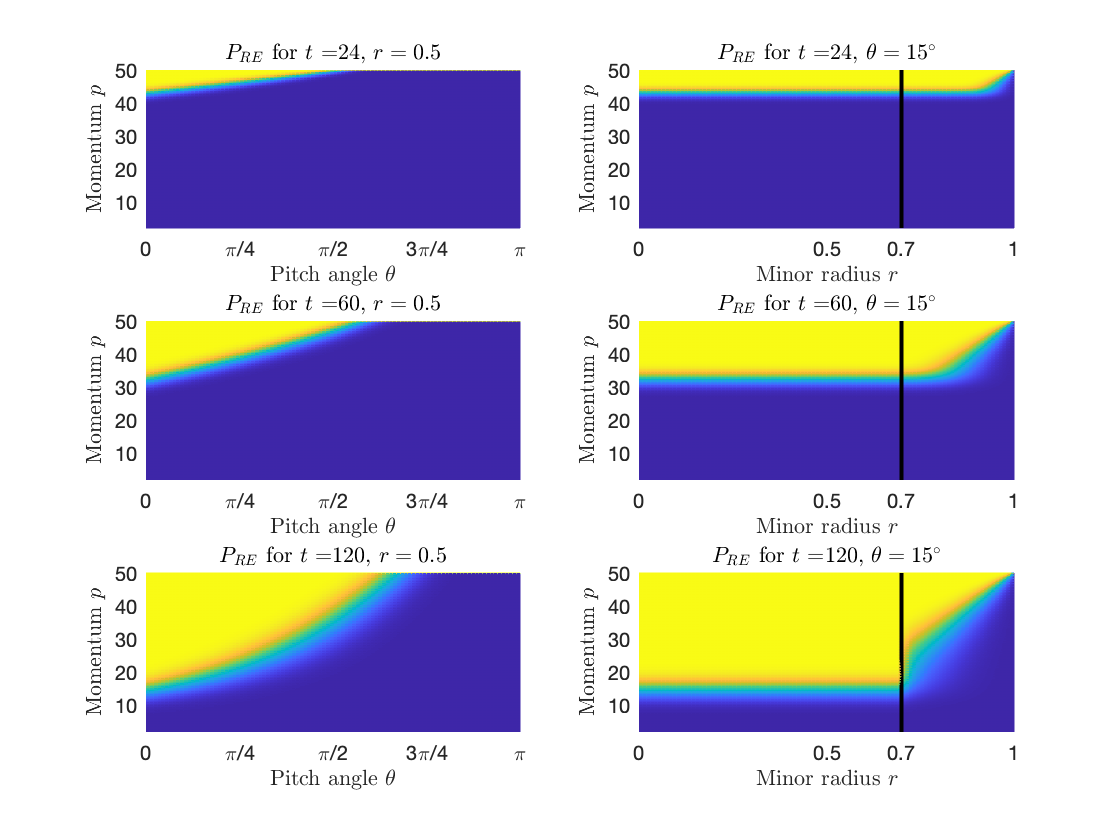}
  \vspace{-0.5cm}
  \caption{
 Same as Fig.~\ref{uniform}, but using a {\em non-uniform grid}. The use of a grid with
a large spacing, $\Delta r_s = 0.1$, for  $r\in[0,r_D]$, and a small spacing, $\Delta r_d = 0.0025$, for $[r_D,1]$,
 eliminates the numerical diffusion and recovers the expected sharp transition between the $P_{RE} \sim 0$ (blue)  and $P_{RE} \sim 1$ (yellow) boundary around $r=r_D=0.67$ (black vertical line).
 The result is in agreement with that shown in Fig.~\ref{denseuni} using a uniform fine grid, but the use of 
 a non-uniform grid significantly reduces the computational cost. 
} 
  \label{nonuni}
\end{figure}

Figures.~\ref{uniform} and \ref{denseuni} show the calculation of the runaway probability using a uniform grid, with blue (yellow) denoting $P_{RE} \sim 0$ ($P_{RE} \sim 1$) values.  Since this is a $3$D problem, the results are displayed in  $(p,\theta)$ planes for different radial cuts, and
in  $(p,r)$ planes for different $\theta$ cuts, at successive times. The relatively small value of the parameter $W$ in the function $f(r)$ introduces a sharp transition in the radial dependence of $V$ and $R$ at $r=r_D$, which creates a boundary layer in the runaway probability.
Resolving this boundary layer using a uniform grid in $r$ requires a significant reduction of the global grid spacing. In particular as shown in Fig.~\ref{uniform} using 
$\Delta r_s=0.1$ gives rise to an artificial spreading of the transition boundary separating the $P_{RE} \sim 1$ and 
$P_{RE} \sim 0$ regions. As shown in Fig.~\ref{denseuni}, to recover the expected sharpness of the transition boundary using a uniform grid requires reducing the global grid spacing to 
$\Delta r_d = 0.0025$. 
However, using a global reduction of the grid spacing is numerically expensive and not needed in the whole computation domain. As an alternative we show in Fig.\ref{nonuni} how the use of a non-uniform grid
can accurately resolve the boundary layer, by using a large spacing, $\Delta r_s = 0.1$, for  $r\in[0,r_D]$ and a small spacing, $\Delta r_d = 0.0025$, for $[r_D,1]$.

\begin{figure}[h!]
\center
  \includegraphics[scale =0.3]{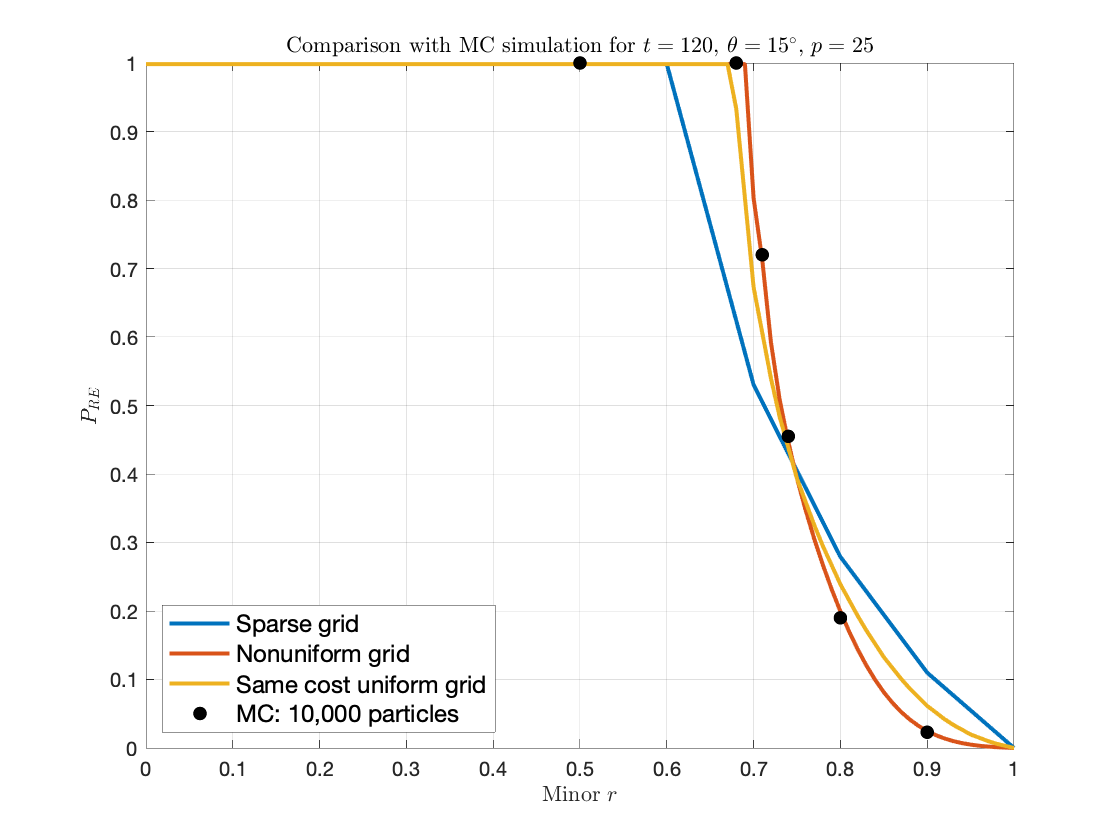}
  \caption{
 Radial dependence of the runaway probability, $P_{RE}$, at $\theta=15^o$ and $p=25$. 
 The black dots correspond to a high resolution direct Monte-Carlo simulation using $10,000$ particles. The
proposed method perfectly recovers in a numerically efficient way this direct expensive reference solution when using a non-uniform grid. Using  a (same cost) uniform grid or a sparse-grid reduces the accuracy of the computation.}
\label{MCplot}
\end{figure}

To further test the accuracy of the proposed method, Fig.~\ref{MCplot} shows the radial dependence of the runaway probability, $P_{RE}$, at $\theta=15^o$ and $p=25$. Since there are not known analytical solutions, we use as reference a high resolution direct Monte-Carlo simulation using $10,000$ particles. As Fig.~\ref{MCplot} shows, the 
proposed method perfectly recovers the reference solution when using a non-uniform grid. Using  a (same cost) uniform grid or a sparse-grid reduces the accuracy of the computation. 

In the previous discussion, the coefficients of the SDE in (\ref{ex32d}) were assumed time independent. However, in the more realistic and physically relevant cases, both, the electric field and the collision operator are time dependent making Eqs.(\ref{ex3}) non-autonomous. 
To illustrate and test the proposed algorithm in this case, we change the constants 
$\bar{v_T}$, $\bar{\nu}_{ee}$, $\delta$, and $E$ in (\ref{parameters}) to the functions 
\begin{equation}
\bar{v_T}(t)=\sqrt{\frac{\hat T}{\tilde{T}}}\, , \qquad \bar{\nu}_{ee}(t)=\left(\frac{\tilde{T}}{\hat T}\right)^{3/2}\, \frac{\ln \hat \Lambda}{\ln \tilde{\Lambda}} \, , \qquad  \delta(t)=\sqrt{\frac{2 \hat T}{m c^2}}
\, , \qquad 
E(t)=E_0\left[ \frac{\hat T_0}{\hat T}\right]^{3/2} \, ,
\end{equation}
where the time dependence enters through the function 
$\hat T (t)=\hat T_f + \left(\hat T_0-\hat T_f\right) e^{-  t/ t_*} $,  $mc^2$ is the rest mass of the electron, 
$\ln  \hat \Lambda = 14.9 -\frac{1}{2} \ln 0.28 + \ln  \hat{T}$, and
$\ln  \tilde \Lambda = 14.9 -\frac{1}{2} \ln 0.28 + \ln  \tilde{T}$.
Physically, this model describes the generation of runaway electrons in dynamics scenarios  of magnetic disruptions in which the plasma thermal quench, described with an exponential cooling model, generates a time dependent electric field described by Ohm's law using a Spitzer conductivity model. The constants  $\hat T_0$ and $\hat T_f$ are the plasma temperature before and after the thermal quench, $t_*$, the cooling rate, $E_0$ the pre-quench electric field, and
 $\tilde T$ a reference temperature scale. As an example we consider the parameter values
 \begin{equation}
\begin{aligned}
\label{parameters2d}
%T_{\max} & = 120, \;\; p_{\min} = 2, \;\; p_{\max} = 50,\;\; Z = 1,\;\;  \tau = 10^5 \\
 & Z = 1,\;\;  \tau = 6000 ,\;\;  E_0 =10^{-3},\;\;
 \hat T_0=3 ,\;\; \hat T_f=0.005,\;\; \tilde T=3
 \\
 & r_D  =  0,\;\;   W = 10^{-4},\;\;  \Delta p = 2,\;\;  V_0 = 0  \, ,
\end{aligned}
\end{equation}
and, to illustrate the dependence on time and radial diffusion, we  
vary the parameters $t_*$ and $D_0$. 

\begin{figure}[h!]
\center
  \includegraphics[scale =0.5]{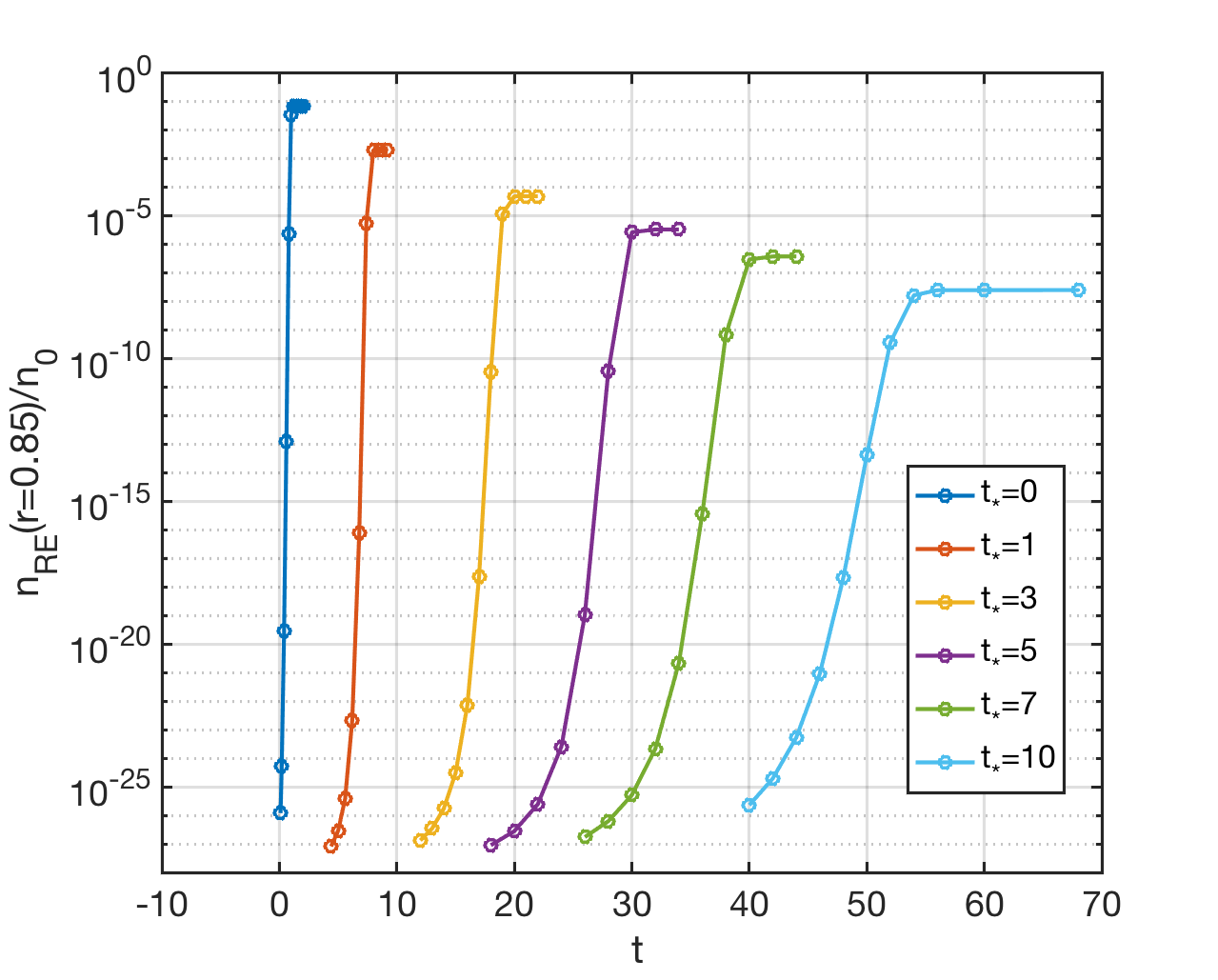}
  \caption{
Dependence of time evolution of seed runaway electron density, $n_{RE}$ at $r_0=0.85$ (normalized to the plasma density, $n_0$) on cooling rate time scale, $t_*$,  for $D_0=0.01$.}
\label{nre_time}
\end{figure}

\begin{figure}[h!]
\center
  \includegraphics[scale =0.5]{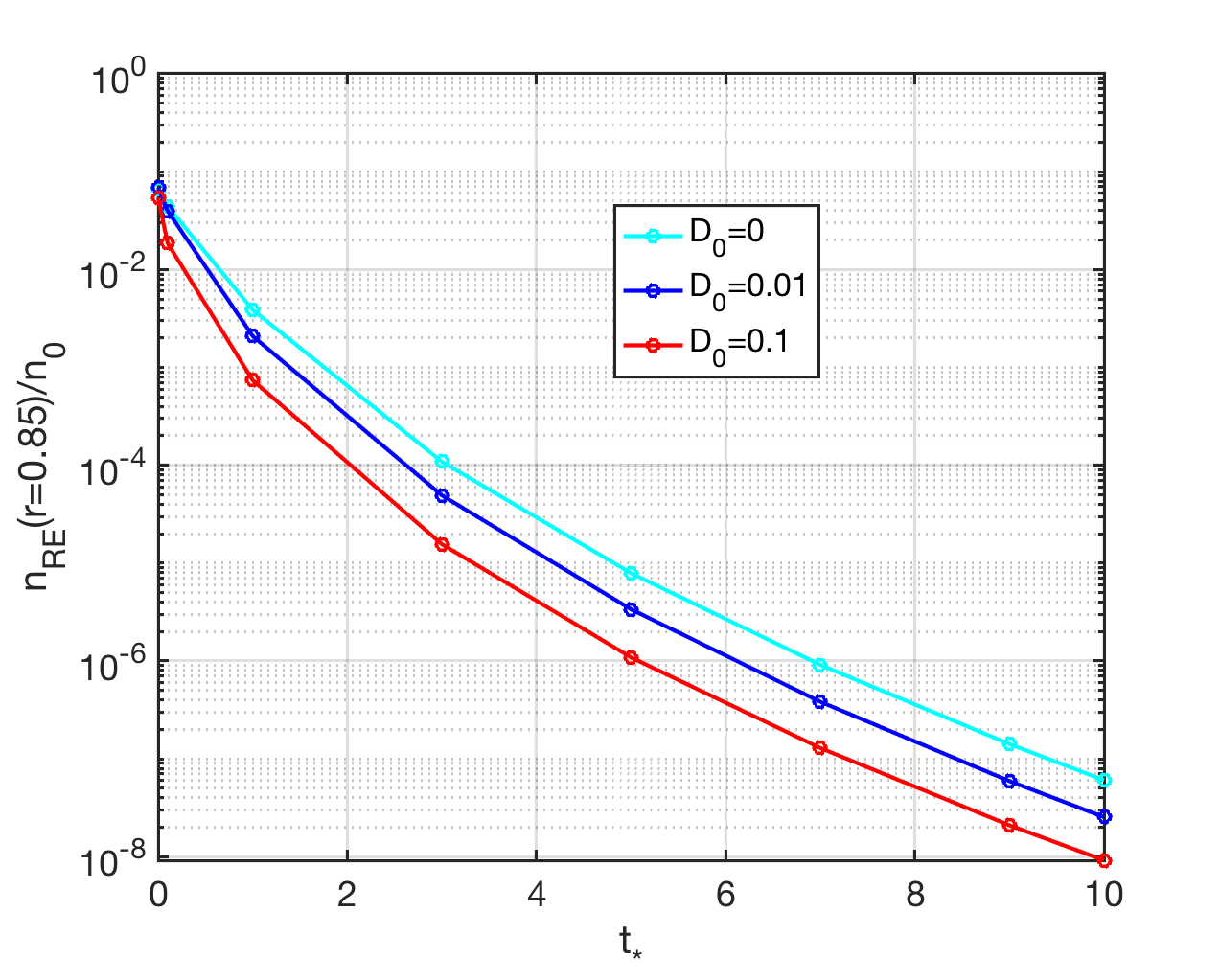}
  \caption{
 Dependence of final seed runaway electron density, $n_{RE}$ at $r_0=0.85$ (normalized to the plasma density, $n_0$) on cooling rate time scale, $t_*$, and radial diffusivity, $D_0$.}
\label{nre_t_star_D}
\end{figure}

Figure \ref{nre_time} shows the time evolution of the density of seed nunaway electrons, $n_{RE}$, at the fixed radius $r=r_0=0.85$ defined as
\begin{equation}
n_{RE}(r_0,t)= 2 \pi n_0\,  \int_{-1}^1 d \xi \int_0^\infty dp\, f_M(p) P_{RE}(\xi, p,r=r_0,t) \, ,
\end{equation}
where $n_0$ is the electron plasma density, $f_M$, is a Maxwellian distribution with temperature $T_0$, and 
$P_{RE}(\xi, p,r=r_0,t)$ is the computed 3D, time dependent probability of runaway at $r=r_0$.
Consistent with the runaway electron physics, it is observed that longer cooling rates (large $t_*$) delay the onset of the  seed runaway electrons production.
In addition, as Fig.~\ref{nre_t_star_D} shows, $t_*$ and $D_0$, have a direct impact on the final saturation level of $n_{RE}$. As expected, fast cooling (small $t_*$) leads to larger final saturation. On the other hand, increasing $D_0$ results in smaller $n_{RE}$ because radial diffusion deconfines a fraction of the electrons before they can be accelerated to runaway energies. 

\section{Concluding Remarks}\label{sec:clu}
% In this work, we proposed a matrix structure probabilistic scheme for efficiently solving the time-independent high-dimensional exit probability based on the relationship between the SDEs and parabolic PDEs. As discussed in Section \ref{prob:set}, the exit probability can be represented by the solution of a parabolic terminal boundary value problem and the proposed scheme utilizes Gaussian quadrature rule and pchip interpolation to compute the integrals involved in the conditional expectation. We write the proposed scheme as a matrix form and use GPU architecture to accelerate the computation for updating the interpolation matrix. Even though our algorithm can handle three-dimensional numerical examples, the full-tensor grids lead to huge computational costs when we extend our approach to higher dimensional problems. Also the refinement sparse grid technique cannot be couple to our algorithm directly. Our future efforts will focus on extending the proposed numerical scheme to the irregular spatial domains and the compatibility of refinement spatial grids.

We have presented an accurate and efficient method for the computation of the exit time probability for a given system of non-autonomous (time-dependent) stochastic differential equations.  The proposed numerical scheme can calculate this probability simultaneously for batches of particles with different initial condition and it is therefore straightforward to parallelize. The theoretical foundation of the proposed method is the Feynman-Kac formula that links the probability of exit time  to the solution of the adjoint Fokker-Planck equation. We express the conditional expectations as integrals evaluated using Gauss-Hermite quadrature rules and pchip interpolation strategies, which can achieve first-order  convergence rate. To compute the full time evolution of the exit time probability by using a single run of the algorithm we formulate the numerical scheme in matrix form. 
 
We presented three examples illustrating specific aspects and advantages of the proposed numerical method. 
The first example considered the exit time probability in a standard 1D Brownian motion and
illustrated the convergence properties of our approach as well as the efficiency of the GPU-accelerated algorithm.
 The second example 
considered the exit time in a 2D advection-diffusion problem with a time dependent incompressible fluid velocity field. This example showed that our numerical scheme is more optimal than the finite difference method regarding stability, accuracy, and computational cost. 
The third example presented a 3D  plasma physics motivated problem and illustrated the seamless incorporation of nonuniform spatial grids in our scheme and the versatility of the method to solve complex physics  problems demanding nontrivial temporal-spatial discretizations. 

Our next step is to extend the proposed method to enable its use in more complex problems involving the coupling of SDEs models with already existing simulators describing parameters in the SDEs and/or external degrees of freedom. For example, in the plasma physics problem discussed in Sec.~\ref{e200} the electric field, $E$, is taken as a fixed parameter. However, in reality, the spatio temporal evolution of $E$ is governed Maxwell's equations, and it  is thus important to extend the method to incorporate the coupling of the SDEs with a PDE solver for $E$. In this plasma physics problem there is also the need to 
go beyond the simplified radial diffusion model used in Sec.~\ref{e200} and 
incorporate advanced models for the spatial evolution of the electrons. 
This task, which is quite challenging in the context of PDE based methods, can be accomplished by replacing the Euler scheme in \eqref{ref-X} with the temporal propagators provided by the external
particle simulator.

% Even though our numerical scheme can handle three-dimensional numerical examples,  full-tensor grids lead to substantial computational costs for higher-dimensional problems. So far, the refinement sparse grid technique cannot be directly coupled to our algorithm. Our future efforts will focus on extending the proposed numerical scheme to the irregular spatial domains and the compatibility of refinement sparse grids.

%\section{Acknowledgments}\label{sec:clu} 

% \vspace{0.5 cm}
\section*{Acknowledgements}
This material is based upon work supported in part by the U.S.~Department of Energy, Office of Science, Offices of Advanced Scientific Computing Research and Fusion Energy Science, and by the Laboratory Directed Research and Development program at the Oak Ridge National Laboratory, which is operated by UT-Battelle, LLC, for the U.S.~Department of Energy under Contract DE-AC05-00OR22725.

\bibliography{library,unpub,escape_ref}
\bibliographystyle{abbrv}

\newpage
\begin{center}
 {\bf Significance and novelty} 
\end{center}

We present an accurate and efficient numerical method for solving the full dynamics of the probability distribution, denoted by $P(t,x)$, of the first exit time of stochastic differential equations with time-dependent parameters. The first exit time is arguably one of the most natural and important transport problems, and this work is motivated by critical physics and engineering applications, i.e., the modeling of runaway acceleration of electrons in realistic plasma physics systems. The main novelty of the proposed method are
\begin{itemize}
    \item[(a)] The unconditional stability for any $\Delta t$, first-order convergence with respect to $\Delta t$, and second-order convergence with respect to $\Delta x$;
    \item[(b)] The output of our method can be used to obtain the exit probability of the SDE with different initial conditions by doing simple convolution;
    \item[(c)] Recovering the entire dynamics of the exit probability $P(t,x)$ by only one temporal iteration, and easy implementation on GPUs. 
\end{itemize}
% \vspace{-0.1cm}
To our knowledge, our method is the first one that possesses all the three features among existing methods for computing $P(t,x)$. In comparison, the MC methods have feature (c); the PDE solvers for the forward Fokker-Planck equations have feature (a), (c); the implicit PDE solvers for the backward (adjoint) Fokker-Planck equations have feature (a), (b); the explicit solvers for the backward (adjoint) Fokker-Planck equations have feature (b), (c).

\end{document}